\documentclass[11pt,a4paper]{article}
\usepackage[margin=22mm]{geometry}
\usepackage[T1]{fontenc}
\usepackage[utf8]{inputenc}
\usepackage{lmodern}
\usepackage{xcolor}
\usepackage{amsmath,amssymb,bm}
\usepackage{graphicx}
\usepackage{longtable,booktabs,array,multirow,calc}
\usepackage{microtype}
\usepackage{cite}
\usepackage{xurl}
\usepackage[hidelinks,hypertexnames=false]{hyperref}
\usepackage{caption}
\usepackage{adjustbox}
\usepackage{placeins}
\usepackage{needspace}
\usepackage{newunicodechar}
\newunicodechar{ǧ}{\u{g}}
\newunicodechar{‐}{-}
\newunicodechar{×}{\ensuremath{\times}}
\allowdisplaybreaks
\title{A Markovian SuASR epidemic framework with stochastic external factors}
\author{Vasileios E. Papageorgiou\textsuperscript{1,2*} \and Georgios Vasiliadis\textsuperscript{2}}
\date{\small\parbox{\textwidth}{\centering
\textsuperscript{1}Department of Mathematics, Aristotle University of Thessaloniki, Thessaloniki, 54124, Greece\\
\textsuperscript{2}Department of Mathematics, University of Western Macedonia, Kastoria, 52100, Greece\\[4pt]
*Correspondence: \href{mailto:vpapageor@math.auth.gr}{vpapageor@math.auth.gr}}}
\begin{document}

\maketitle

\begin{abstract}

Epidemic waves are often shaped by uncertain external drivers, such as
environmental variability or behavioral and policy changes, that can
randomly amplify or dampen transmission. Capturing this randomness
requires models whose dynamics evolve under time-varying, stochastic
influences. In this paper, we introduce a stochastic
Susceptible--Asymptomatic--Symptomatic--Removed (SuASR) framework that
embeds random external factors directly into the epidemic mechanism
through an extended continuous-time Markov chain (CTMC). The proposed
framework is flexible enough to represent epidemic dynamics both with
reinfection and with infection-induced immunity, depending on the
epidemiological setting under study. Unlike standard Markovian epidemic
models, our formulation explicitly tracks two concurrent infectious
sources (asymptomatic and symptomatic), yielding coupled infection
pathways and substantially richer transient behavior. A key contribution
is analytical tractability, where we derive computable expressions for
the joint distribution of asymptomatic and symptomatic infection-event
counts and develop tools tailored to extreme-outbreak risk, including
the distribution of the maximum infectious burden. These results enable
probabilistic quantification of epidemic severity beyond mean
trajectories, providing direct estimates of the likelihood of rare but
consequential surges. We validate the model through extensive
sensitivity analysis and an application to the H1N1 influenza epidemic,
illustrating how stochastic external influences reshape outbreak
profiles, total infections from both sources, and the probability of
extreme epidemic scenarios.

\end{abstract}



\section{Introduction}

The modeling of infectious diseases based on advanced mathematical
models have flourished during the last decades were more advanced
deterministic and stochastic approaches are proposed. Deterministic
methods, utilizing non-linear systems of differential equations, like
the SIR model, aim to predict epidemic trajectories efficiently
\cite{ref1,ref2,ref3}. Each letter corresponds to one of the model's compartments
as the population is assumed to be partitioned in three classes/states,
namely the susceptible (S), infectious (I) and recovered (R). Several
extensions have been proposed which account for the existence of exposed
individuals \cite{ref4,ref5}, deaths \cite{ref6,ref7}, vaccinations \cite{ref8}, or
additional control measures \cite{ref9,ref10,ref11}. However, the aforementioned
deterministic approaches struggle to capture the time-varying dynamics
of epidemics, while the stability analysis of epidemic equilibria based
on compartmental schemes with constant parameters is not always
informative.

In contrast, stochastic modeling is useful for capturing the effects of
demographic, transmission, and environmental variability \cite{ref12,ref13,ref14}.
These methods often predict epidemic spread using Bayesian estimation
techniques, such as Kalman and particle filters \cite{ref15,ref16,ref17,ref18,ref19,ref20,ref21,ref22,ref23}. Other
studies focus on stochastic descriptors, offering insights about the
progression of the phenomenon. Examples include time- and size-related
features such as the maximum number of infectious cases \cite{ref24}, the
time of infection or mortality \cite{ref12,ref25}, the disease extinction
duration \cite{ref26}, the quasi-stationarity of the system \cite{ref24,ref27,ref28},
etc. Some research delves into vaccine-focused approaches, determining
optimal revaccination times or quantifying total infectious cases
\cite{ref29,ref30,ref31}.

Moreover, related stochastic epidemic models have already addressed
several final-outcome quantities. Ball and Clancy studied final size and
severity in stochastic multitype epidemic models \cite{ref32}, while Ball
and O'Neill derived exact results for final size and more general
final-state random variables \cite{ref33}. Other works have considered
symptomatic and asymptomatic infectives, Markovian random environments,
and computational methods for final-outcome or final-size distributions
\cite{ref27,ref34,ref35}. More recently, symptomatic and asymptomatic infectious
classes have also been studied under semi-Markov switching \cite{ref36}.

It is known that there are many epidemics which are characterized by
varying parameters and more specifically seasonal changes, like measles,
diphtheria, chickenpox, cholera or influenza-type viruses \cite{ref37}.
These changes may arise from environmental factors, such as temperature
or humidity, or from the implementation of public health measures.
Moreover, the variability of parameters, particularly the timing of
transitions between environmental states, is marked by uncertainty, as
the emergence of a new variant or the impact of a vaccination campaign
or lockdown cannot be predicted with absolute precision. Given this,
deterministic methods may be inadequate. Therefore, the development of
new stochastic approaches that can capture both parameter variability
and uncertainty is essential.

In this paper, we introduce a stochastic
\textbf{Su}sceptible-\textbf{A}symptomatic-\textbf{S}ymptomatic-\textbf{R}emoved
(SuASR) epidemic model which considers the existence of random external
factors, based on an extended CTMC. Random factors may be influenced by
social patterns, climatic variables, or the establishment of control
measures like vaccination or lockdowns which can significantly alter the
parameters that dictate the prevalence of the infectious disease
\cite{ref38,ref39}. Existing literature mainly focuses on models where
environmental changes are either not taken into account or are assumed
to be deterministic. For instance, the prevalent assumption involves
disease transmission following a sinusoidal function, indicating that
the contact rate fluctuates over time but without considering
uncertainty \cite{ref40,ref41,ref42,ref43}. However, there is limited attention given to
models incorporating stochastic variations in the environment.
Indicatively, we refer to the papers \cite{ref27,ref44} and \cite{ref45} where the
authors explore epidemic models with random environment, employ a
cyclically alternating renewal process according to the specificities of
the SIR scheme and introduce a deterministic SIS model with Markovian
environment, correspondingly.

Compared to the aforementioned studies, our model accounts for two
sources of infection (arising from both asymptomatic and symptomatic
cases) thereby capturing more complex and realistic epidemic dynamics.
The proposed framework can accommodate epidemic dynamics both with
reinfection (Variant I) and with infection-induced immunity (Variant
II), according to the epidemiological context under consideration.

Additionally, we derive formulas for stochastic descriptors that are
relevant to epidemic severity, focusing on both cumulative infection
burden and extreme transmission events. In particular, we study the
joint distribution of symptomatic and asymptomatic infection-event
counts, as well as the distribution of the peak number of infectious
individuals, using a suitably modified Markov chain. Thus, our
contribution is the matrix-recursive computation of these distributional
quantities for a finite-population SuASR model whose parameters vary
according to external factor states.

Efficient computational schemes are introduced to compute joint and
marginal distributions, along with their moments. A comprehensive
extensive sensitivity analysis is conducted to explore the model's
behavior in relation to key parameters. To further demonstrate the
credibility and practical value of the proposed stochastic framework, we
present a parameter-based scenario analysis motivated by the 2009 H1N1
influenza pandemic in Mexico. Finally, we offer insights into efficient
computational strategies for evaluating the proposed stochastic
descriptors.

Overall, the main novel contributions of this analysis are: a) a new
Markovian epidemic model (SuASR) incorporating both symptomatic and
asymptomatic cases, b) a stochastic process with time-varying parameters
accounting for random external factors, c) formulas for efficiently
computing the joint distribution of symptomatic and asymptomatic cases,
d) expressions for estimating the maximum number of simultaneously
infectious cases, e) demonstration of the model's behavior through
sensitivity analysis and a parameter-based H1N1 scenario analysis.

\section{Methodology}

We introduce a novel stochastic epidemiological model that accounts for
the existence of random external factors. These factors may relate to
environmental elements such as temperature or humidity, emergence of new
disease strains, public health interventions, and more. Incorporating
these random factors stands as a primary innovation in our analysis,
allowing for the accommodation of inherent variations within natural
phenomena. Initially, we outline a deterministic compartmental framework
before transitioning to its stochastic counterpart. After that, a series of important epidemiological descriptors is
provided, examining the joint distribution of symptomatic and
asymptomatic cases and the maximum infectious burden.

\subsection{The stochastic SuASR framework with random external factors}

We introduce a stochastic SuASR framework with random external factors
through an extended CTMC. To accommodate different epidemiological
settings, we consider two related formulations that share the same
stochastic architecture but differ in the way recovery is modeled. The
two variants share the same state space architecture and differ only in
the way individuals are routed after infection. We first present the
recurrent-infection formulation and then describe the
immunity-preserving formulation used when reinfections are not expected
within the outbreak horizon.

\subsubsection{Variant I: recurrent-infection formulation}

Before delving into the stochastic epidemic model featuring random
external factors, it is important to outline the foundational
epidemiological framework upon which we build the stochastic equivalent
model. The deterministic SuASR framework consists of a system of four
ordinary differential equations (ODEs), describing the evolution in the
number of susceptible, asymptomatic, symptomatic, and removed cases. The
corresponding compartmental structure is illustrated in Equation (1),
where

\begin{equation}
\begin{aligned}
&\frac{dS(t)}{dt} = - \frac{\beta}{N}S(t)\left( I(t) + A(t) \right) + \gamma I(t) + \kappa A(t) , \\
&\frac{dA(t)}{dt} = (1 - p)\frac{\beta}{N}S(t)\left( I(t) + A(t) \right) - \kappa A(t) , \\
&\frac{dI(t)}{dt} = p\frac{\beta}{N}S(t)\left( I(t) + A(t) \right) - \gamma I(t) - \mu I(t) , \\
&\frac{dR(t)}{dt} = \mu I(t) ,
\end{aligned}\tag{1}\label{eq:1}
\end{equation}

for \(t \geq 0.\) The term \(N\) denotes the population size, \(\beta\)
the disease's transmission/infection rate, \(p\) the ratio of
symptomatic infections, \(\mu\) the removal rate and \(\gamma\) and
\(\kappa\) the recovery rates from a symptomatic or an asymptomatic
infection, respectively. Moreover, \(S(t)\), \(I(t)\), \(A(t)\) and
\(R(t)\) represent the number of susceptible, symptomatic, asymptomatic
and removed cases at time \(t\).

Recovery from asymptomatic or symptomatic infection returns individuals
to the susceptible class, so infection does not necessarily confer
immunity in this formulation. Since the rate \(\mu\) may be associated
with disease-induced removal of symptomatic individuals, we include the
corresponding transition only from the symptomatic compartment to the
removed class. This choice is justified by the (almost) negligible
mortality rate among asymptomatic cases, as the absence of symptoms
typically indicates less severe infections. Moreover, the population
under study is assumed to be closed, with homogeneous contact patterns
among individuals.

Now, for the transition to the stochastic point of view, we employ a
CTMC, namely
\(\mathcal{X} =\left\{ \bm{X}(t) = \left( I(t),A(t),R(t) \right), t \geq 0 \right\}.\)
The fourth state of system (1) can be described as a linear combination
of the three remaining states, i.e. \(S(t) = N - I(t) - A(t) - R(t)\),
for \(t \geq 0\). Considering that the Markov chain is in the
initial/current state \((i,a,r)\) at the time \(t\), the transition
rates of the Markov chain are given by expression (2), where

\begin{equation}
u_{(i,a,r),( i^{'},a^{'},\ r^{'} )} = \begin{cases}
- u_{i,a,r},   & ( i^{'}, a^{'}, r^{'} ) = (i,a,r), \\
p\left( \frac{\beta}{N} \right)_{i,a,r},   & ( i^{'}, a^{'}, r^{'} ) = (i + 1,a,r), \\
(1 -  p)\left( \frac{\beta}{N} \right)_{i,a,r},   & ( i^{'}, a^{'}, r^{'} ) = (i,a + 1,r), \\
\gamma_{i},   & ( i^{'}, a^{'}, r^{'} ) = (i - 1,a,r), \\
\kappa_{a},   & ( i^{'}, a^{'}, r^{'} ) = (i,a - 1,r), \\
{  \mu}_{i},   & ( i^{'}, a^{'}, r^{'} ) = (i - 1,a,r + 1), \\
0,   & \text{otherwise},

\end{cases}\tag{2}\label{eq:2}
\end{equation}

while
\(p\left( \frac{\beta}{N} \right)_{i,a,r} = p\left( \frac{\beta}{N} \right)(i + a)(N - i - a - r)\),
\((1 - p)\left( \frac{\beta}{N} \right)_{i,a,r} = (1 - p)(\frac{\beta}{N})(i + a)(N - i - a - r)\),
\(\gamma_{i} = \gamma i\), \(\kappa_{a} = \kappa a\),
\(\mu_{i} = \mu i\) and
\(u_{i,a,r} = p\left( \frac{\beta}{N} \right)(i + a)(N - i - a - r) + (1 - p)\left( \frac{\beta}{N} \right)(i + a)(N - i - a - r) + \gamma i + \kappa a + \mu i\).

The form of Equation (2) derives from the fact that we have five
possible transitions assuming that we initiate the examination of the
phenomenon from the transient state \((i,a,r)\). Ultimately, the Markov
chain \(\mathcal{X}\) transitions from state \((i,a,r)\), to the state

\begin{enumerate}
\def\labelenumi{\roman{enumi})}
\item
  \((i,a + 1,r),\) with rate
  (\(1 - p)\left( \frac{\beta}{N} \right)_{i,a,r}\), when a susceptible
  case becomes asymptomatic,
\item
  \((i + 1,a,r),\) with rate
  \(p\left( \frac{\beta}{N} \right)_{i,a,r}\), when a susceptible case
  becomes symptomatic,
\item
  \((i - 1,a,r),\) with rate \(\gamma_{i}\), when a symptomatic case
  recovers,
\item
  \((i,a - 1,r),\) with rate \(\kappa_{a}\), when an asymptomatic case
  recovers,
\item
  \((i - 1,a,r + 1),\) with rate \(\mu_{i}\), when a symptomatic
  individual is removed.
\end{enumerate}

The infinitesimal generator matrix
\(\bm{U} = (u_{(i,a,r),( i^{'}, a^{'}, r^{'} )}:\ (i,a,r), ( i^{'}, a^{'}, r^{'} ) \in A_{U})\),
is a square matrix having a block tridiagonal form. The corresponding
state space
\(A_{U} = \{(i,a,r):0 \leq i \leq N,\ 0 \leq a \leq N - i,\ 0 \leq r \leq N - i - a\}\)
contains a total of
\(\left| A_{U} \right| = \sum_{i = 0}^{N}{\sum_{a = 0}^{N - i}{(N - i - a + 1) = \frac{(N + 1)(N + 2)(N + 3)}{6}}}\)
states.

As we mentioned in the Introduction, one of the main objectives of the
present analysis is the development of an epidemic model that accounts
for external random factors, which can alter the parameters that
determine the progression of the disease. This mechanism is governed by
the CTMC \(\mathcal{F} = \{F(t),t \geq 0\}\) with state space
\(A_{F} = \{ f : 1 \leq f \leq F\}\) and transition rates
\(u_{F}(f,f')\), for \(1 \leq f,f^{'} \leq F\). This construction
represents external changes, such as environmental variation or
externally imposed policy regimes, rather than interventions triggered
by the current number of infectious individuals. If the total exit rate
from an external factor state (EFS) is positive, then the holding time
is exponentially distributed with mean equal to the reciprocal of the
sum of all transition rates leaving that state. If the total exit rate
is zero, then EFS is absorbing for the external-factor process.

We propose an extension of the Markov chain \(\mathcal{X}\) by adding a
fourth component, which is associated with a finite number of (EFSs). We
introduce the Markov chain
\(\widetilde{\mathcal{X}} = \{\widetilde{\bm{X}}(t) = \left( I(t),A(t),R(t),F(t) \right), t \geq 0\}\).
An extension of the state space is also dictated, where
\(A = A_{U} \times A_{F} = \left\{ (i,a,r,f):0 \leq i \leq N,\ 0 \leq a \leq N - i,\ 0 \leq r \leq N - i - a,\ 1 \leq f \leq F \right\}\).
Thus, the cardinality of the extended state space is
\(|A| = \left| A_{F} \right| \cdot \left| A_{U} \right| = F\frac{(N + 1)(N + 2)(N + 3)}{6}\).
In parallel, a new infinitesimal generator matrix
\(\widetilde{\bm{U}} = \{{\widetilde{u}}_{(i,a,r,f),( i^{'},a^{'},r^{'},f^{'} )}: (i,a,r,f),( i^{'},a^{'},r^{'},f^{'} ) \in A\}\)
is constructed, where

\begin{equation}
{\widetilde{u}}_{(i,a,r,f),( i^{'},a^{'},r^{'},f^{'} )} = \begin{cases}
- {\widetilde{u}}_{i,a,r,f},   & ( i^{'},a^{'},r^{'},f^{'} ) = (i,a,r,f), \\
p(f)\left( \frac{\beta(f)}{N} \right)_{i,a,r},   & ( i^{'},a^{'},r^{'},f^{'} ) = (i + 1,a,r,f), \\
\left( 1 -  p(f) \right)\left( \frac{\beta(f)}{N} \right)_{i,a,r}, & ( i^{'},a^{'},r^{'},f^{'} ) = (i,a + 1,r,f), \\
\gamma_{i}(f),   & ( i^{'},a^{'},r^{'},f^{'} ) = (i - 1,a,r,f), \\
\kappa_{a}(f),   & ( i^{'},a^{'},r^{'},f^{'} ) = (i,a - 1,r,f), \\
{  \mu}_{i}(f),   & ( i^{'},a^{'},r^{'},f^{'} ) = (i - 1,a,r + 1,f), \\
u_{F}( f,f^{'}),   & ( i^{'},a^{'},r^{'},f^{'} ) = (i,a,r,f'), f \neq f^{'}, \\
0,   & \text{otherwise},  

\end{cases}\tag{3}\label{eq:3}
\end{equation}

while
\({\widetilde{u}}_{i,a,r,f} = p(f)\left( \frac{\beta(f)}{N} \right)_{i,a,r} + \left( 1 - \ p(f) \right)\left( \frac{\beta(f)}{N} \right)_{i,a,r} + \gamma_{i}(f) + \kappa_{a}(f){+ \mu}_{i}(f) + \sum_{f^{'} = 1}^{F}{u_{F}(f,f^{'})}\),
for \(f \neq f^{'} \in A_F.\) Term \(u_{F}\left( f,f^{'} \right)\) denotes
the transition rate from external factor state (EFS) \(f\) to
\(f' \neq f \in A_F\).

The state space \(A\) is divided into three nested levels, namely
\(A = \bigcup_{i = 0}^{N}{l(i)}\), where the \(l(i)\) represents the
first sublevel which is associated with \(I(t) = i.\) Furthermore,
\(l(i) = \bigcup_{a = 0}^{N - i}{l(i,a)}\) and
\(l(i,a) = \bigcup_{r = 0}^{N - i - a}{l(i,a,r)}\), where
\(l(i,a,r) = \left\{ (i,a,r,f):1 \leq f \leq F \right\}\). The sets
\(l(i,a)\) and \(l(i,a,r)\) represent the second and third level,
correspondingly. The respective cardinalities are
\(\left| l(i,a,r) \right| = F\),
\(\left| l(i,a) \right| = F(N - i - a + 1)\) and
\(\left| l(i) \right| = \sum_{a = 0}^{N - i}{F(N - i - a + 1)} = F\frac{(N - i + 1)(N - i + 2)}{2}\).
Finally, according to the aforementioned structure, \(l(0,0)\)
represents a closed set of disease-free states as there are no
symptomatic or asymptomatic individuals who can spread the virus or
trigger a secondary infection wave. Therefore, when the process enters
the set \(l(0,0)\) the examination of the phenomenon terminates. This
formulation is intended for epidemic settings in which repeated
infections may occur over the time horizon of interest, and it is the
version used in the sensitivity analyses of Sections 3.1 and 3.2.

\subsubsection{Variant II: immunity-preserving formulation}

As noted in the Introduction, the removed state may include individuals
who are either deceased or have acquired immunity following infection.
For epidemic settings in which infected individuals develop complete
immunity, or in which immunity may be imperfect but the observation
window is too short for reinfections to occur, we introduce a modified
version of the Markov chain defined in Equation (2). In this case, the
infinitesimal generator matrix takes the following form

\begin{equation*}
u_{(i,a,r),( i^{'}, a^{'}, r^{'})} = \begin{cases}
- u_{i,a,r},   & ( i^{'}, a^{'}, r^{'}) = (i,a,r), \\
p\left( \frac{\beta}{N} \right)_{i,a,r},   & ( i^{'}, a^{'}, r^{'}) = (i + 1,a,r), \\
(1 -  p)\left( \frac{\beta}{N} \right)_{i,a,r},   & ( i^{'}, a^{'}, r^{'}) = (i,a + 1,r), \\
\kappa_{a},   & ( i^{'}, a^{'}, r^{'}) = (i,a - 1,r + 1), \\
{  (\gamma + \mu)}_{i},   & ( i^{'}, a^{'}, r^{'}) = (i - 1,a,r + 1), \\
0,   & \text{otherwise}.  

\end{cases}
\end{equation*}

The key modification, relative to Equation (2), lies in the recovery
transitions. Infected individuals now move only to the removed class
rather than returning to the susceptible class. The extension of the
process to incorporate random external factors through the addition of a
fourth component can be carried out exactly as in Section 2.1.1. Under
this formulation, fewer infections are expected, since recovered
individuals enter the removed state and no longer participate in disease
transmission.

For the remainder of the paper, the recursive forms developed in
Sections 2.2 and 2.3 remain valid for Variant II after replacing the
generator and all associated submatrices by their immunity-preserving
counterparts. In particular, the recovery transitions in the relevant
generator blocks now lead to the removed class rather than to the
susceptible class. In the compact matrix formulation, this affects the
matrices in Equation (20), while for the maximum infectious burden it
affects the transition structure in Equation (23). This
immunity-preserving formulation is employed in Section 3.3 for the H1N1
application.

\subsection{The number of symptomatic and asymptomatic infections}

At this point, we aim to construct formulas for the computation of the
joint distribution of the number of symptomatic
\(\left( N^{S} \right)\ \)and asymptomatic \(\left( N^{A} \right)\)
infections until the eradication of the disease (total number). Thus, we
examine the aforementioned stochastic descriptor until the Markov chain
enters the set \(l(0,0)\). Let
\(T = \inf\left\{ t \geq 0\ :\left( I(t),A(t),R(t),F(t) \right) = (0,0,r,f), 0 \leq r \leq N, 1 \leq f \leq F \right\}\)
be the random variable which is associated with the extinction of the
disease due to the absence of active infectious cases.

For both variants, the cumulative number of infection events until
extinction is finite under the parameter conditions considered in this
paper. Indeed, the CTMC has a finite epidemic state space and the
disease-free set is absorbing. Moreover, when the recovery and removal
rates are positive, no closed communicating class exists outside the
disease-free set. Hence, absorption into the disease-free set occurs
with probability one, and the extinction time is finite almost surely.
In Variant I, the same individual may be infected more than once, so the
cumulative infection-event counts are not necessarily bounded by \(N\),
but they remain finite almost surely.

The following theorem describes the computation of the joint
distribution of the number of symptomatic/asymptomatic infections,
denoted by \({\widetilde{p}}_{i,a,r,f}(n,m)\), for
\((n,m) \in \left( \mathbb{N}\cup\left\{ 0 \right\} \right)^{2}\) and
\((i,a,r,f) \in A\). Specifically, \({\widetilde{p}}_{i,a,r,f}(n,m)\)
denotes the probability that \(n\) new symptomatic and \(m\)
asymptomatic cases will occur until disease's extinction. For the
purposes of the next theorem, we assemble the corresponding
probabilities into the column vector
\({\widetilde{\bm{p}}}_{i,a}(n,m)\), where \(i\), \(a\), \(n\) and
\(m\) are considered fixed. The subscripts \(i\) and \(a\) refer to the
initial numbers of symptomatic and asymptomatic infections,
respectively. It should be noted that \(n\) and \(m\) refer to the
number of infections occurring from the time at which the analysis
begins until the extinction of the disease, excluding the initial number
of symptomatic and asymptomatic cases.

\textbf{Theorem 1.} The joint distribution of the number of
symptomatic/asymptomatic cases until the disease's extinction can be
computed through Equations (4)--(17). For \(n = m = 0\), we have
\begin{equation}
{\widetilde{\bm{p}}}_{0,0}(0,0) = \bm{1}_{\left| l(0,0) \right|},\tag{4}\label{eq:4}
\end{equation}
while
\begin{equation}
{\widetilde{\bm{p}}}_{0,0}(n,m) = \bm{0}_{\left| l(0,0) \right|},\quad (n,\ m) \neq (0,0),\tag{5}\label{eq:5}
\end{equation}
when there are no infected (symptomatic or asymptomatic) cases to the
initial state. Now, when \(i \neq 0\) or \(a \neq 0\), we obtain
\begin{equation}
{\widetilde{\bm{p}}}_{i,0}(0,0) = - {\widetilde{\bm{U}}}_{(i,0),(i,0)}^{- 1}{\widetilde{\bm{U}}}_{(i,0),(i - 1,0)}{\widetilde{\bm{p}}}_{i - 1,0}(0,0),\quad 1 \leq i \leq N,\tag{6}\label{eq:6}
\end{equation}
\begin{equation}
{\widetilde{\bm{p}}}_{0,a}(0,0) = - {\widetilde{\bm{U}}}_{(0,a),(0,a)}^{- 1}{\widetilde{\bm{U}}}_{(0,a),(0,a - 1)}{\widetilde{\bm{p}}}_{0,a - 1}(0,0),\quad 1 \leq a \leq N,\tag{7}\label{eq:7}
\end{equation}
while when both indices are non-zero, we obtain
\begin{equation}
\begin{aligned}
&\widetilde{\bm{p}}_{i,a}(0,0) = -\widetilde{\bm{U}}_{(i,a),(i,a)}^{-1}\Bigl[\
\widetilde{\bm{U}}_{(i,a),(i-1,a)}\widetilde{\bm{p}}_{i-1,a}(0,0) 
+ \widetilde{\bm{U}}_{(i,a),(i,a-1)}\widetilde{\bm{p}}_{i,a-1}(0,0)\Bigr],  1\le i\le N-1,\\
& 1\le a\le N-i.
\end{aligned}
\tag{8}\label{eq:8}
\end{equation}
Also, when \(n > 0,\)
\begin{equation}
\begin{aligned}
&\widetilde{\bm{p}}_{i,0}(n,0)=-\widetilde{\bm{U}}_{(i,0),(i,0)}^{-1}\Bigl[
\bigl(1-\delta_{N-i,0}\bigr)\widetilde{\bm{U}}_{(i,0),(i+1,0)}\widetilde{\bm{p}}_{i+1,0}(n-1,0) 
+ \widetilde{\bm{U}}_{(i,0),(i-1,0)}\widetilde{\bm{p}}_{i-1,0}(n,0)\Bigr], \\
&1\le i\le N,
\end{aligned}
\tag{9}\label{eq:9}
\end{equation}
\begin{equation}
\begin{aligned}
&\widetilde{\bm{p}}_{0,a}(n,0) = -\widetilde{\bm{U}}_{(0,a),(0,a)}^{-1}\Bigl[ 
\bigl(1-\delta_{N-a,0}\bigr)\widetilde{\bm{U}}_{(0,a),(1,a)}\widetilde{\bm{p}}_{1,a}(n-1,0)+ \widetilde{\bm{U}}_{(0,a),(0,a-1)}\widetilde{\bm{p}}_{0,a-1}(n,0)\Bigr],  \\
& 1\le a\le N,
\end{aligned}
\tag{10}\label{eq:10}
\end{equation}
and
\begin{equation}
\begin{aligned}
\widetilde{\bm{p}}_{i,a}(n,0) &= -\widetilde{\bm{U}}_{(i,a),(i,a)}^{-1}\Bigl[ \bigl(1-\delta_{N-a-i,0}\bigr)\widetilde{\bm{U}}_{(i,a),(i+1,a)}\widetilde{\bm{p}}_{i+1,a}(n-1,0)+ \widetilde{\bm{U}}_{(i,a),(i-1,a)}\widetilde{\bm{p}}_{i-1,a}(n,0) \\
&+ \widetilde{\bm{U}}_{(i,a),(i,a-1)}\widetilde{\bm{p}}_{i,a-1}(n,0)\Bigr], 1\le i\le N-1, 1\le a\le N-i.
\end{aligned}
\tag{11}\label{eq:11}
\end{equation}
On the other hand, when \(m > 0\ \)and \(n = 0\), we obtain
\begin{equation}
\begin{aligned}
&\widetilde{\bm{p}}_{i,0}(0,m) = -\widetilde{\bm{U}}_{(i,0),(i,0)}^{-1}\Bigl[ \bigl(1-\delta_{N-i,0}\bigr)\widetilde{\bm{U}}_{(i,0),(i,1)}\widetilde{\bm{p}}_{i,1}(0,m-1) + \widetilde{\bm{U}}_{(i,0),(i-1,0)}\widetilde{\bm{p}}_{i-1,0}(0,m)\Bigr],\\
&1\le i\le N,
\end{aligned}
\tag{12}\label{eq:12}
\end{equation}
\begin{equation}
\begin{aligned}
&\widetilde{\bm{p}}_{0,a}(0,m) = -\widetilde{\bm{U}}_{(0,a),(0,a)}^{-1}\Bigl[ \bigl(1-\delta_{N-a,0}\bigr)\widetilde{\bm{U}}_{(0,a),(0,a+1)}\widetilde{\bm{p}}_{0,a+1}(0,m-1) + \widetilde{\bm{U}}_{(0,a),(0,a-1)}\widetilde{\bm{p}}_{0,a-1}(0,m)\Bigr], \\
& 1\le a\le N,
\end{aligned}
\tag{13}\label{eq:13}
\end{equation}
and
\begin{equation}
\begin{aligned}
\widetilde{\bm{p}}_{i,a}(0,m) &= -\widetilde{\bm{U}}_{(i,a),(i,a)}^{-1}\Bigl[ \bigl(1-\delta_{N-a-i,0}\bigr)\widetilde{\bm{U}}_{(i,a),(i,a+1)}\widetilde{\bm{p}}_{i,a+1}(0,m-1) + \widetilde{\bm{U}}_{(i,a),(i-1,a)}\widetilde{\bm{p}}_{i-1,a}(0,m) \\
&+ \widetilde{\bm{U}}_{(i,a),(i,a-1)}\widetilde{\bm{p}}_{i,a-1}(0,m)\Bigr], 1\le i\le N-1, 1\le a\le N-i.
\end{aligned}
\tag{14}\label{eq:14}
\end{equation}
Finally, when \(m > 0\) and \(n > 0\), we obtain
\begin{equation}
\begin{aligned}
\widetilde{\bm{p}}_{i,0}(n,m) &= -\widetilde{\bm{U}}_{(i,0),(i,0)}^{-1}\Bigl[ \bigl(1-\delta_{N-i,0}\bigr)\widetilde{\bm{U}}_{(i,0),(i+1,0)}\widetilde{\bm{p}}_{i+1,0}(n-1,m)  \\
&+ \bigl(1-\delta_{N-i,0}\bigr)\widetilde{\bm{U}}_{(i,0),(i,1)}\widetilde{\bm{p}}_{i,1}(n,m-1)+ \widetilde{\bm{U}}_{(i,0),(i-1,0)}\widetilde{\bm{p}}_{i-1,0}(n,m)\Bigr], 1\le i\le N,
\end{aligned}
\tag{15}\label{eq:15}
\end{equation}
\begin{equation}
\begin{aligned}
\widetilde{\bm{p}}_{0,a}(n,m) &= -\widetilde{\bm{U}}_{(0,a),(0,a)}^{-1}\Bigl[ \bigl(1-\delta_{N-a,0}\bigr)\widetilde{\bm{U}}_{(0,a),(1,a)}\widetilde{\bm{p}}_{1,a}(n-1,m) \\
&+ \bigl(1-\delta_{N-a,0}\bigr)\widetilde{\bm{U}}_{(0,a),(0,a+1)}\widetilde{\bm{p}}_{0,a+1}(n,m-1) + \widetilde{\bm{U}}_{(0,a),(0,a-1)}\widetilde{\bm{p}}_{0,a-1}(n,m)\Bigr], 1\le a\le N,
\end{aligned}
\tag{16}\label{eq:16}
\end{equation}
and
\begin{equation}
\begin{aligned}
\widetilde{\bm{p}}_{i,a}(n,m) &= -\widetilde{\bm{U}}_{(i,a),(i,a)}^{-1}\Bigl[ \bigl(1-\delta_{N-a-i,0}\bigr)\widetilde{\bm{U}}_{(i,a),(i+1,a)}\widetilde{\bm{p}}_{i+1,a}(n-1,m) \\
&+ \bigl(1-\delta_{N-a-i,0}\bigr)\widetilde{\bm{U}}_{(i,a),(i,a+1)}\widetilde{\bm{p}}_{i,a+1}(n,m-1) + \widetilde{\bm{U}}_{(i,a),(i-1,a)}\widetilde{\bm{p}}_{i-1,a}(n,m) \\
&+ \widetilde{\bm{U}}_{(i,a),(i,a-1)}\widetilde{\bm{p}}_{i,a-1}(n,m)\Bigr], 1\le i\le N-1, 1\le a\le N-i.
\end{aligned}
\tag{17}\label{eq:17}
\end{equation}

Submatrices
\({\widetilde{\bm{U}}}_{(i,a),( i^{'},a^{'} )}\) include
the transition rates that are associated with transitions from states
with \(i\) symptomatic and \(a\) asymptomatic to states with \(i^{'}\)
symptomatic and \(a^{'}\) asymptomatic cases.

\textbf{Proof of Theorem 1.} Conditioning on the event that determines
the next transition out of the current state
\((i,a,r,f) \in A_{T} = A - l(0,0)\), we get

\begin{equation}
\begin{aligned}
{\widetilde{u}}_{i,a,r,f}\ {\widetilde{p}}_{i,a,r,f}(n,m) &= p(f)\left( \frac{\beta(f)}{N} \right)_{i,a,r}{\widetilde{p}}_{i + 1,a,r,f}(n - 1,m)\left( 1 - \delta_{N - i - a - r,0} \right)\left( 1 - \delta_{n0} \right)\ \\
&+ (1 - p(f))\left( \frac{\beta(f)}{N} \right)_{i,a,r}{\widetilde{p}}_{i,a + 1,r,f}(n,m - 1)\left( 1 - \delta_{N - i - a - r,0} \right)(1 - \delta_{m0}) \\
& + \mu_{i}(f){\widetilde{p}}_{i - 1,a,r + 1,f}(n,m)\left( 1 - \delta_{i0} \right) + \gamma_{i}(f){\widetilde{p}}_{i - 1,a,r,f}(n,m)\left( 1 - \delta_{i0} \right) \\
&+ \kappa_{a}(f){\widetilde{p}}_{i,a - 1,r,f}(n,m)\left( 1 - \delta_{a0} \right) + \sum_{\substack{f'=1\\f'\ne f}}^F u_{F}(f,f^{'})\widetilde{p}_{i,a,r,f'}(n,m) ,
\end{aligned}\tag{18}\label{eq:18}
\end{equation}

for the set of transient states, while quantities \((1 - \delta_{kl})\)
are used to eliminate probability terms without physical meaning. We
note that \(\delta_{kl}\) denotes the Kronecker function, which equals
one when \(k = l\); otherwise, \(\delta_{kl} = 0\). Now, formula (18)
can be described in matrix form based on formulas (6)--(17). Employing
Equations (6)--(17) we can compute the joint distribution of the total
number of new symptomatic/asymptomatic infections.

The last term of (18) describes the transitions between the EFSs, where
\(f' \neq f \in \{ 1,\ldots,F\}\). The model's parameters are related to
the current EFS and are characterized by time-varying behavior. Finally,
the probability vectors
\({\widetilde{\bm{p}}}_{0,0}(0,0) = \bm{1}_{\left| l(0,0) \right|}\)
and
\({\widetilde{\bm{p}}}_{0,0}(n,m) = \bm{0}_{\left| l(0,0) \right|}\ \)for
\((n,m) \neq (0,0)\), are associated with the set of absorbing states
\(l(0,0)\). Apparently, the absence of asymptomatic and symptomatic
cases renders the generation of new infections infeasible. \(\square\)

\textbf{Remark.} According to Equation (18) we can introduce an
alternative expression for the computation of the joint distribution.
More specifically, we obtain

\begin{equation}
\widetilde{\bm{p}}(n,m) = - \left( {\widetilde{\bm{U}}}_{T}^{*} \right)^{- 1}\left( \left( 1 - \delta_{n0} \right){\widetilde{\bm{U}}}_{T}^{+}\widetilde{\bm{p}}(n - 1,m) + \left( 1 - \delta_{m0} \right){\widetilde{\bm{U}}}_{T}^{-}\widetilde{\bm{p}}(n,m - 1) \right),\tag{19}\label{eq:19}
\end{equation}

for
\((n,m) \in ({\mathbb{N}\cup\{ 0\})}^{2} - \left\{ (0,0) \right\},\)
where
\({\widetilde{\bm{U}}}_{T}^{*} = ({\widetilde{u}}_{(i,a,r,f),( i^{'},a^{'},r^{'},f^{'} )}^{*}:\ (i,a,r,f),( i^{'},a^{'},r^{'},f^{'} ) \in A_{T})\),
with

\begin{equation}
{\widetilde{u}}_{(i,a,r,f),( i^{'},a^{'},r^{'},f^{'} )}^{*} = \begin{cases}
- {\widetilde{u}}_{i,a,r,f},   & ( i^{'},a^{'},r^{'},f^{'} ) = (i,a,r,f), \\
\gamma_{i}(f),   & ( i^{'},a^{'},r^{'},f^{'} ) = (i - 1,a,r,f), \\
\kappa_{a}(f),   & ( i^{'},a^{'},r^{'},f^{'} ) = (i,a - 1,r,f), \\
\mu_{i}(f),   & ( i^{'},a^{'},r^{'},f^{'} ) = (i - 1,a,r + 1,f), \\
u_{F}( f,f^{'} ),   & ( i^{'},a^{'},r^{'},f^{'} ) = ( i,a,r,f^{'} ),  f^{'} \neq f, \\
0,   & \text{otherwise},  

\end{cases}\tag{20}\label{eq:20}
\end{equation}

while
\({\widetilde{\bm{U}}}_{T}^{+} = \left({\widetilde{u}}_{(i,a,r,f),( i^{'},a^{'},r^{'},f^{'} )}^{+}:\ (i,a,r,f),( i^{'},a^{'},r^{'},f^{'} ) \in A_{T}\right)\),
with

\begin{equation}
{\widetilde{u}}_{(i,a,r,f),( i^{'},a^{'},r^{'},f^{'} )}^{+} = \begin{cases}
p(f)\left( \frac{\beta(f)}{N} \right)_{i,a,r}   & ( i^{'},a^{'},r^{'},f^{'} ) = (i + 1,a,r,f), \\
0   & \text{otherwise},  

\end{cases}\tag{21}\label{eq:21}
\end{equation}

and
\({\widetilde{\bm{U}}}_{T}^{-} = \left({\widetilde{u}}_{(i,a,r,f),( i^{'},a^{'},r^{'},f^{'} )}^{-}:\ (i,a,r,f),( i^{'},a^{'},r^{'},f^{'} ) \in A_{T}\right)\),
with

\begin{equation}
{\widetilde{u}}_{(i,a,r,f),( i^{'},a^{'},r^{'},f^{'} )}^{-} = \begin{cases}
\left( 1 - p(f) \right)\left( \frac{\beta(f)}{N} \right)_{i,a,r} & ( i^{'}, a^{'}, r^{'},f^{'} ) = (i,a + 1,r,f), \\
0   & \text{otherwise}
.
\end{cases}\tag{22}\label{eq:22}
\end{equation}

The 3-dimensional tensor \(\widetilde{\bm{p}}(n,m)\) contains the
joint probabilities for every initial state of the set \(A_{T}\). Since
\(\widetilde{\bm{p}}(0,0)\) is known (Equations (4)--(8)), the
recursive computation of the joint distribution becomes feasible. An
alternative derivation of the \(\widetilde{\bm{p}}(0,0)\) vector
using generating functions is provided in Appendix A.

Formula (19) seems simpler compared to the matrix equations of Theorem
1. However, it requires the construction and storage of matrices of
higher dimension. On the other hand, based on (4)--(17) we obtain a more
efficient computational scheme, while we generate only the necessary
matrices for each initial state\(.\) Therefore, we recommend the
utilization of (19) and (4)--(17) for small and larger population sizes,
respectively. Finally, an algorithmic procedure for the computation of
the \(\widetilde{\bm{p}}(n,m)\) tensor is presented in Appendix A.

Consequently, the cumulative quantities \(N^{S}\) and \(N^{A}\) should
be interpreted as symptomatic and asymptomatic infection-event counts.
In Variant I, since recovered individuals return to susceptibility, the
same individual may contribute more than once to these counts.
Therefore, \(N^{S}\) and \(N^{A}\) are not necessarily bounded by the
population size (\(N\)).

\subsection{Distribution of the maximum number of simultaneously infectious individuals}

In this section, we are interested in investigating extreme tendencies
of the system, such as the maximum number of infectious
(symptomatic/asymptomatic) cases until the disease's eradication. To
facilitate this attempt, we reconstruct the Markov chain of Section 2.2.
Namely, we define the absorbing Markov chain
\(\mathcal{Y} =\left\{ \bm{Y}(t) = \left( I(t) + A(t),A(t),R(t),F(t) \right), t \geq 0 \right\}\),
considering the existence of random external factors that may
differentiate the epidemic's parameters.

The new Markov process is associated with the introduction of a new
state space, namely
\(B = \left\{ (i + a,a,r,f):\quad 0 \leq i + a \leq N,\quad 0 \leq a \leq i + a,\quad 0 \leq r \leq N - i - a,\quad 1 \leq f \leq F \right\}.\)
The new state space can be partitioned in sublevels, too, where
\(B = \bigcup_{i + a = 0}^{N}{l'(i + a)}\ \)and
\(l'(i + a) = \bigcup_{a = 0}^{i + a}{l'(i + a,a)}.\) Moreover,
\(l'(i + a,a) = \{(i + a,a,r,f):(r,f) \in \{ 0,\ldots,N - a - i\} \times \{ 1,\ldots,F\}\}\)
leading to the absorbing set of states
\(l'(0,0) = \{(0,0,r,f): (r,f) \in \{ 0,\ldots,N\} \times \{ 1,\ldots,F\}\}\),
and the set of transient states \(\bigcup_{i + a = 1}^{N}{l'(i + a)}.\) For the analysis of the maximum infectious burden, we stop the process upon its first entry into the disease-free set \(l'(0)\).
Accordingly, the transition rates in Equation \eqref{eq:23} apply only to states
with \(i+a>0\); all transition rates out of disease-free states are
set to zero.

The respective infinitesimal generator matrix
\(\widetilde{\bm{Q}} = \{{\widetilde{q}}_{(i + a,a,r,f),( i^{'} + a^{'},a^{'},r^{'},f^{'} )}: (i + a,a,r,f),( i^{'} + a^{'},a^{'},r^{'},f^{'} ) \in B\}\),
has the form

\begin{equation}
\begin{adjustbox}{max width=0.94\linewidth}
$\displaystyle {\widetilde{q}}_{(i + a,a,r,f),( i^{'} + a^{'},a^{'},r^{'},f^{'} )} = \begin{cases}
- {\widetilde{q}}_{i + a,a,r,f},   & ( i^{'} + a^{'}, a^{'}, r^{'},f^{'} ) = (i + a,a,r,f), \\
p(f)\left( \frac{\beta(f)}{N} \right)_{i,a,r},   & ( i^{'} + a^{'}, a^{'}, r^{'},f^{'} ) = (i + a + 1,a,r,f), \\
\left( 1 -  p(f) \right)\left( \frac{\beta(f)}{N} \right)_{i,a,r},   & ( i^{'} + a^{'}, a^{'}, r^{'},f^{'} ) = (i + a + 1,a + 1,r,f), \\
\gamma_{i}(f),   & ( i^{'} + a^{'}, a^{'}, r^{'},f^{'} ) = (i + a - 1,a,r,f), \\
\kappa_{a}(f),   & ( i^{'} + a^{'}, a^{'}, r^{'},f^{'} ) = (i + a - 1,a - 1,r,f), \\
{  \mu}_{i}(f),   & ( i^{'} + a^{'}, a^{'}, r^{'},f^{'} ) = (i + a - 1,a,r + 1,f), \\
u_{F}( f,f^{'}),   & ( i^{'} + a^{'}, a^{'}, r^{'},f^{'} ) = ( i + a,a,r,f^{'} ), f^{'} \neq f, \\
0,   & \text{otherwise},  

\end{cases}$
\end{adjustbox}
\tag{23}\label{eq:23}
\end{equation}

where \(\tilde{q}_{i+a,a,r,f}=\tilde{u}_{i,a,r,f}\), for \(i+a>0\), while

\begin{equation}
\widetilde{\bm{Q}} = \begin{pmatrix}
\bm{0}_{\left| l^{'}(0) \right| \times \left| l^{'}(0) \right|} & \bm{0}_{\left| l^{'}(0) \right| \times \left| \bigcup_{i + a = 1}^{N}{l^{'}(i + a)} \right|} \\
{\widetilde{\bm{Q}}}_{0} & {\widetilde{\bm{Q}}}_{T}
\end{pmatrix},\tag{24}\label{eq:24}
\end{equation}

where matrix \({\widetilde{\bm{Q}}}_{T}\) contains the transition
rates between the transient states and \({\widetilde{\bm{Q}}}_{0}\)
connects the transient with the absorbing states. Their forms are

\begin{equation}
{\widetilde{\bm{Q}}}_{T} = \begin{pmatrix}
\begin{matrix}
{\widetilde{\bm{Q}}}_{1,1} & {\widetilde{\bm{Q}}}_{1,2} \\
{\widetilde{\bm{Q}}}_{2,1} & {\widetilde{\bm{Q}}}_{2,2}
\end{matrix} & \cdots & \begin{matrix}
\bm{0} \\
\bm{0}
\end{matrix} \\
 \vdots & \ddots & \vdots \\
\begin{matrix}
\bm{0\quad } & \bm{0}
\end{matrix} & \cdots & {\widetilde{\bm{Q}}}_{N,N}
\end{pmatrix}\ ,\tag{25}\label{eq:25}
\end{equation}

and

\begin{equation}
{\widetilde{\bm{Q}}}_{0} = \begin{pmatrix}
{\widetilde{\bm{Q}}}_{1,0} \\
\bm{0}_{\left| \bigcup_{i + a = 2}^{N}{l^{'}(i + a)} \right| \times |l^{'}(0)|}
\end{pmatrix}.\tag{26}\label{eq:26}
\end{equation}

Matrices \({\widetilde{\bm{Q}}}_{i,j}\) contain the transition rates
from level \(l'(i)\) (\(i\) infectious cases) to \(l'(j)\) (\(j\)
infectious case).

Since we are interested in the maximum number of simultaneously
infectious cases until the disease's extinction, we define the random
variable
\[
\tau=\inf\bigl\{t\geq0:\ (I(t)+A(t),A(t),R(t),F(t))=(0,0,r,f), 0\leq r\leq N,1\leq f\leq F\bigr\},
\]
which represents the duration of the epidemic. Next, we define the
random variable \(\xi = \max_{t \in \lbrack 0,\tau)}{(I(t) + A(t))}\)
that denotes the maximum number of simultaneously infectious cases. Our
objective is the computation of the cumulative distribution function of
the maximum number of simultaneously infectious cases,

\begin{equation}
F_{i+a,a,r,f}(x)=\begin{cases}
0, & x<1,\\
P\bigl(\xi\le x\mid\bm{Y}(0)=(i+a,a,r,f)\bigr), & 1\le x<N,\\
1, & x\ge N.
\end{cases}
\tag{27}\label{eq:27}
\end{equation}

It should be noted that the subscripts of the \(F_{i + a,a,r,f}(x)\)
function denote the selected initial state.

\textbf{Theorem 2.} The vector of cumulative distribution functions
which are associated with the maximum number of simultaneously
infectious (symptomatic/asymptomatic) cases, for \(y = 1,\ldots,N\), can
be computed according to

\begin{equation}
\bm{F}(y) = \  - {\widetilde{\bm{Q}}}_{y}^{- 1}\left( \left( \delta_{i + a,1}\delta_{a,1}\kappa_{1}(f) + {\delta_{i + a,1}\delta}_{a,0}\left( \mu_{1}(f) + \gamma_{1}(f) \right) \right)\bm{1}_{|\bigcup_{i + a = 1}^{y}{l^{'}(i + a)}|} \right)\ .\tag{28}\label{eq:28}
\end{equation}

The column vector
\({\widetilde{\bm{q}}}_{0}(y) = \left( \delta_{i + a,1}\delta_{a,1}\kappa_{1}(f) + \delta_{i + a,1}\delta_{a,0}\left( \mu_{1}(f) + \gamma_{1}(f) \right) \right)\bm{1}_{\left| \bigcup_{i + a = 1}^{y}{l^{'}(i + a)} \right|}\)
represents the transition rates from the set of transient states
\(\bigcup_{i + a = 1}^{y}{l'(i + a)}\) to the absorbing set \(l'(0)\).
Matrix \({\widetilde{\bm{Q}}}_{y}\) is a submatrix of
\(\widetilde{\bm{Q}}\) containing the transition rates between the
states of the set \(\bigcup_{i + a = 1}^{y}{l'(i + a)}\).

\textbf{Proof.} For a fixed \(y = 1,\ldots,N\), we notice that
\(F_{i + a,a,r,f}(y)\) equals the conditional probability that starting
from
\(\bm{Y}(0) = \left( I(0) + A(0),A(0),R(0),F(0) \right) = (i + a,a,r,f)\),
the Markov chain \(\mathcal{Y}\) enters the set of states \(l'(0)\)
without visiting the states of \(l'(y + 1).\) The absorption of the
Markov process which derives from the eradication of the disease is
inevitable, since the state space \(B\) contains a unique closed
absorbing set. After removing the rows and columns of matrix
\({\widetilde{\bm{Q}}}_{T}\) that correspond to position where
\(I(t) + A(t) > y\), we ensure that during the outbreak, the maximum
number of simultaneous infectious subjects is at most \(y\). According
to the above, the computation of the vector of cumulative probabilities
\(\bm{F}(y)\) becomes feasible.

Since we are interested in estimating the probability of
\(I(t) + A(t) \leq y\), for a fixed \(y = 1,\ \ldots,\ N\), we introduce
the auxiliary Markov chain
\(\mathcal{Y}(y) = \{ \bm{Y}(y,t),\quad t \geq 0\}\). This chain is defined in
the state space
\(B(y) = l'(0)\bigcup_{}^{}{\left( \bigcup_{i + a = 1}^{y}{l'(i + a)} \right)\bigcup_{}^{}{l'(y + 1)}}\),
where \(l'(0)\) and \(l'(y + 1)\) represent sets of absorbing states.
For simplicity, these sets can be lumped into two absorbing states,
leading to
\(B(y) = \{ 0\}\bigcup_{}^{}{\left( \bigcup_{i + a = 1}^{y}{l'(i + a)} \right)\bigcup_{}^{}{\{ y + 1}\}}\).

Finally, the corresponding generator matrix becomes \cite{ref46,ref47}

\begin{equation}
\widetilde{\bm{Q}}(y) = \begin{pmatrix}
0 & \bm{0}_{\left| \bigcup_{i + a = 1}^{y}{l^{'}(i + a)} \right|}^{T} & 0 \\
{\widetilde{\bm{q}}}_{0}(y) & {\widetilde{\bm{Q}}}_{y} & {\widetilde{\bm{q}}}_{y + 1}(y) \\
0 & \bm{0}_{\left| \bigcup_{i + a = 1}^{y}{l^{'}(i + a)} \right|}^{T} & 0
\end{pmatrix},\tag{29}\label{eq:29}
\end{equation}

where \({\widetilde{\bm{q}}}_{0}(y)\) and
\({\widetilde{\bm{q}}}_{y + 1}(y)\) are column vectors of order
\(\bigcup_{i + a = 1}^{y}{l'(i + a)}\). Square matrix
\({\widetilde{\bm{Q}}}_{y}\) is of the same dimension, while it
includes the transition rates between the transient states of the
auxiliary Markov chain.

The column vector
\({\widetilde{\bm{q}}}_{0}(y) = \left( \delta_{i + a,1}\delta_{a,1}\kappa_{1}(f) + \delta_{i + a,1}\delta_{a,0}\left( \mu_{1}(f) + \gamma_{1}(f) \right) \right)\bm{1}_{|\bigcup_{i + a = 1}^{y}{l^{'}(i + a)}|}\),
contains the transition rates from the set
\(\bigcup_{i + a = 1}^{y}{l'(i + a)}\) to the absorbing set \(\{ 0\}\).
The values of parameters \(\mu_{1}(f)\), \(\gamma_{1}(f)\) and
\(\kappa_{1}(f)\) vary according to the EFS associated with each initial
state in \(\bigcup_{i + a = 1}^{y}{l'(i + a)}\). The Kronecker delta
function \(\delta_{a,1}\ \)is used to clarify that the \(\kappa_{1}(f)\)
terms are included in the rows of \({\widetilde{\bm{q}}}_{0}(y)\)
corresponding to the presence of one asymptomatic case. Based on the
above, the vector of conditional probabilities can be obtained using
(28), where
\(\bm{F}(y) = \  - {\widetilde{\bm{Q}}}_{y}^{- 1}{\widetilde{\bm{q}}}_{0}(y)\),
which signifies the probability that \(\mathcal{Y}(y)\) enters
\(\{ 0\}\), without visiting the states of the set \(\{ y + 1\}\).
Finally,

\begin{equation}
{\widetilde{\bm{q}}}_{y + 1}(y) = \  - {\widetilde{\bm{Q}}}_{y}\bm{1}_{\left| \bigcup_{i + a = 1}^{y}{l^{'}(i + a)} \right|} - {\widetilde{\bm{q}}}_{0}(y),\tag{30}\label{eq:30}
\end{equation}

as the rows of \(\widetilde{\bm{Q}}(y)\) sum to zero. \(\square\)

Equation (28) is also intuitively consistent, as for a fixed
\(y = 1,\ldots,\ N\), matrix \(- {\widetilde{\bm{Q}}}_{y}^{- 1}\)
provides the expected time that the process \(\mathcal{Y}(y)\), spends
in the transient states times a column vector that contains the
transition rates to the set \(l'(0)\). Consequently, the process does
not visit any states where the number of simultaneously infectious cases
exceeds the selected threshold \(y.\)

It should be noted that the infection force per time step,
\(I(t) + A(t)\), assumes that symptomatic and asymptomatic infectious
individuals make the same per capita contribution to transmission. Thus,
the infection rate per external factor, \(\beta(f), f = 1,\ldots,F\),
is interpreted as a common effective transmission rate for both
infectious classes. This simplifying assumption keeps the
finite-population distributional analysis tractable and is also used in
the H1N1 illustration of Section 3.3, because separate reliable
estimates of symptomatic and asymptomatic transmissibility are not
available.

\section{Results}

In this section, we examine how different sets of parameters influence
the behavior of the stochastic SuASR framework under random external
factors and the analyzed descriptors. By systematically exploring
multiple parameter sets under diverse initial and environmental factors,
we aim to derive analytical insights into the stochastic behavior of
epidemic spread. These findings can inform public health decision-making
by enabling the formulation of more targeted and temporally responsive
intervention strategies.

For the remainder of this section, we employ relatively small and
moderate population sizes for illustrative purposes. Thus, we use
formula (19) to compute the joint distribution of new symptomatic and
asymptomatic infections. However, the use of moderate population size is
appropriate for several localized epidemic contexts, including outbreaks
in hospital wards, schools, military training camps, and other closed or
semi-closed communities, where stochastic effects remain non-negligible
and finite-population modeling is justified \cite{ref25,ref48}. It is important
to stress that \(N\) does not necessarily represent the population of an
entire geographical region, but rather the subset of individuals who are
realistically involved in the transmission process. Under this
interpretation, the initial number of susceptible individuals becomes an
epidemiologically meaningful parameter that may itself require
estimation, as discussed in \cite{ref12,ref49,ref50,ref51}. Therefore, moderate
population sizes may also be suitable when analyzing epidemic spread on
a broader scale.

Sections 3.1 and 3.2 are based on Variant I and are intended as methodological sensitivity analyses. 
For the numerical evaluation of the joint infection-count distributions in these sections, we use the truncation limits
\(n_{\max}=m_{\max}=150\), retaining counts \(0\leq n,m\leq150\).
Section 3.3 uses Variant II for the
H1N1 influenza application, where reinfections are assumed negligible
over the study period.

\subsection{The impact of system's time-varying parameters}

In this section, we investigate the impact of model parameters on the
joint distribution of the number of symptomatic/asymptomatic infections,
along with the maximum number of simultaneously infectious cases, using
the model variant presented in Section 2.1.1. We assume that the state
space of external random effects consists of three states, i.e.
\(f \in \{ 1,\ 2,\ 3\}\). For illustrative purposes, we consider a
population size of \(N\  = \ 20\), and symptomatic ratio
\(p\  = \ 0.6\). Higher population sizes can be examined in a similar
manner. Moreover, we select
\(\bm{\beta} \in \{(0.2,\ 0.7,\ 1.2),\ (0.6,\ 1.1,\ 1.6)\}\),
\(\bm{\gamma} =\bm{\kappa} \in \{(0.3,\ 0.15,\ 0.05),\ (0.5,\ 0.35,\ 0.2)\}\),
and \(\bm{\mu} \in \{(0.02,\ 0.05,\ 0.1),\ (0.05,\ 0.1,\ 0.15)\}\).
These parametric triplets represent a progression towards a more severe
disease, potentially attributed to the onset of variants. We choose
\(u_{F}(1,2) = \frac{1}{90}\), \(u_{F}(2,3) = \frac{1}{60}\), and
\(u_{F}(f,f') = \ 0\), otherwise. The initial state is set as
\((i,\ a,\ r,\ f)\  = (5,\ 5,\ 5,\ 1)\). Based on the above, the process
remains in the first EFS on average for 90 time units, transitions to
the second EFS where it stays for an average of 60 time units and then
persists indefinitely in the third EFS.

The resulting joint probability distributions are illustrated in Figure
1. 
The omitted probability mass is less than \(10^{-6}\) in all instances.
 We observe a higher incidence of symptomatic
infections, as anticipated given \(p = 0.6\). The configuration
\(\bm{\beta} = (0.6,\ 1.1,\ 1.6)\),
\(\bm{\gamma} = \bm{\kappa} = (0.3,\ 0.15,\ 0.05)\), and
\(\bm{\mu} = (0.02,\ 0.05,\ 0.1)\) corresponds to a more severe
manifestation of the disease. Under this scenario, a bimodal
distribution emerges with expected numbers of \(E(N^{S}) = 61.85\) and
\(E(N^{A}) = 41.25\). This pattern results from the intensified
infection rates, which lead to a rapid depletion of susceptible
individuals. At the same time, the relatively lower recovery and
mortality rates prolong the duration of that infectious cases remain at
this phase, thereby facilitating the continued generation of new cases.

Conversely, when considering parameter combinations that hinder the
transmission of the disease, characterized by higher recovery and
mortality and lower transmission rates, we observe unimodal joint
distributions with a peak at zero. Hence, it becomes evident that higher
expected numbers of infections are associated with
greater\(\ \bm{\beta}\ \)and lower \(\bm{\gamma}\),
\(\bm{\kappa}\), and \(\bm{\mu}\) rates. This trend results in a
system with a higher prevalence of infectious cases, serving as the
driving force behind the observed phenomenon.

Now, we choose the initial state
\((i + a,\ a,\ r,\ f)\  = \ (1,\ 1,\ 0,\ 1)\), which corresponds to the
presence of only one asymptomatic case. Unimodal decreasing forms with a
mode at 1 are observed in case that the impact of the exit rates from
the infectious phase surpasses the virus's transmission (Figure 2) which
can lead to the eradication of the disease before the number of
simultaneously infectious individuals exceeds the initial count. In
scenarios more favorable to disease transmission, a second mode appears,
resulting in higher expected values. This pattern is similarly reflected
in the variance, where greater variability (and thus greater
uncertainty) corresponds to a wider range of trajectories for the
stochastic process.

\begin{figure}[p]
\centering
\includegraphics[width=0.32\linewidth]{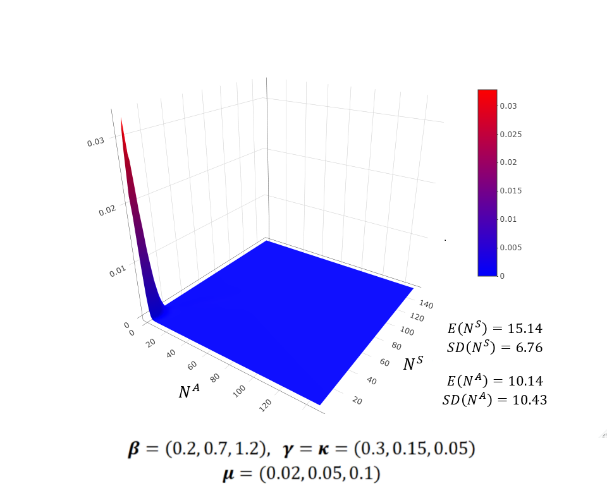}\hfill
\includegraphics[width=0.32\linewidth]{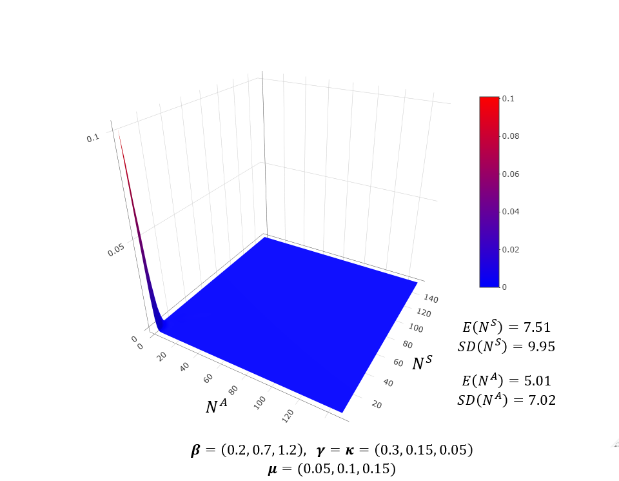}\hfill
\includegraphics[width=0.32\linewidth]{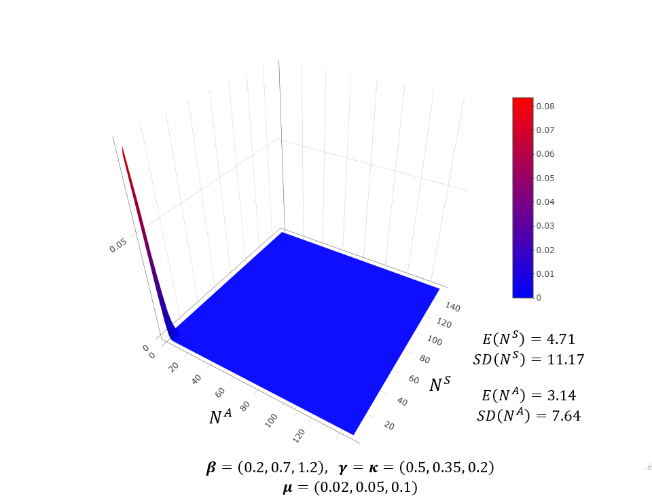}\par\medskip
\includegraphics[width=0.32\linewidth]{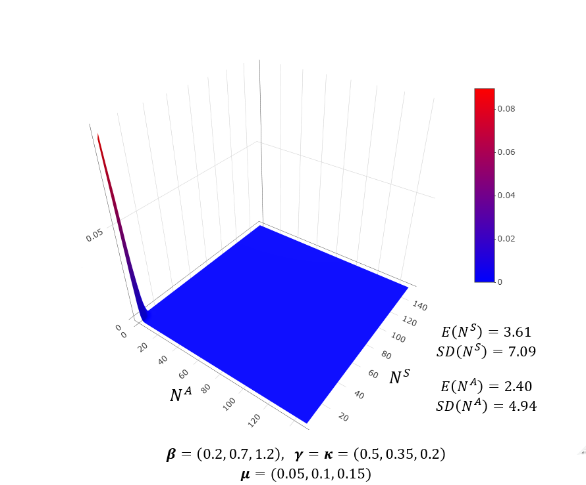}\hfill
\includegraphics[width=0.32\linewidth]{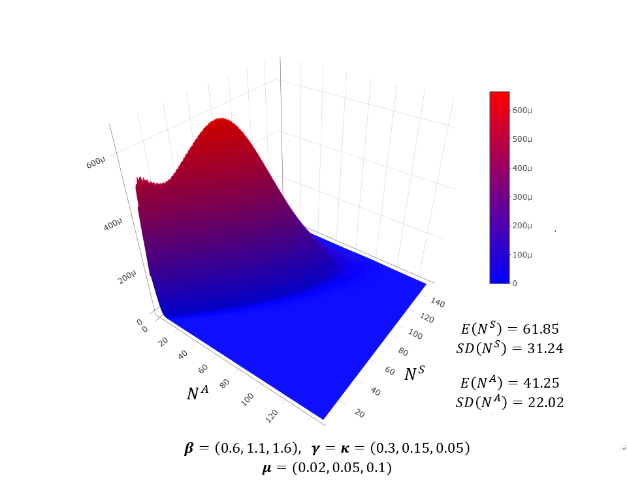}\hfill
\includegraphics[width=0.32\linewidth]{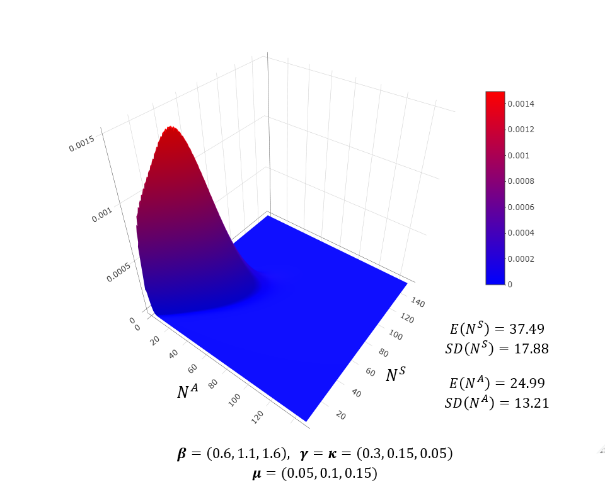}\par\medskip
\includegraphics[width=0.32\linewidth]{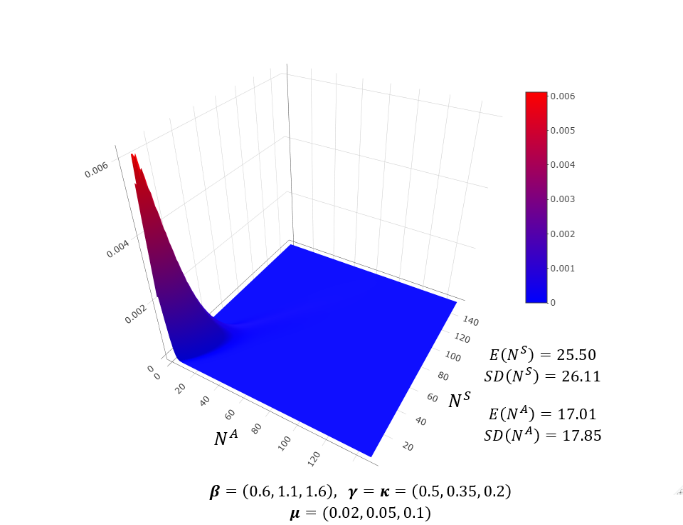}
\includegraphics[width=0.32\linewidth]{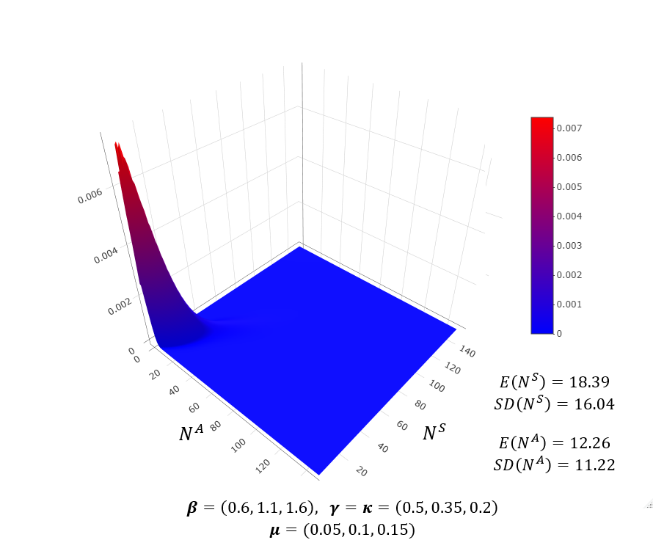}
\caption{Joint probability distributions of the total number
of new symptomatic \(\left( N^{S} \right)\) and asymptomatic \((N^{A})\)
infections. In each graph we include the expected value and standard
deviation of the respective marginal distribution}\label{fig:1}
\end{figure}

\begin{figure}[p]
\centering
\includegraphics[width=0.32\linewidth]{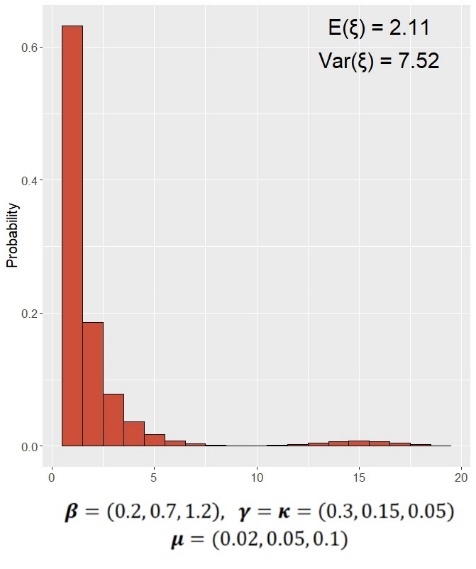}\hfill
\includegraphics[width=0.32\linewidth]{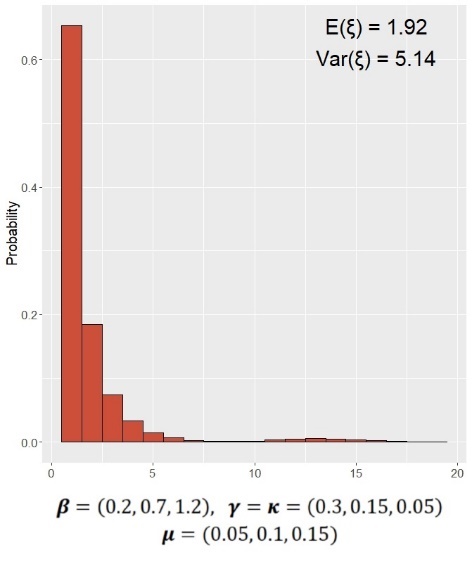}\hfill
\includegraphics[width=0.32\linewidth]{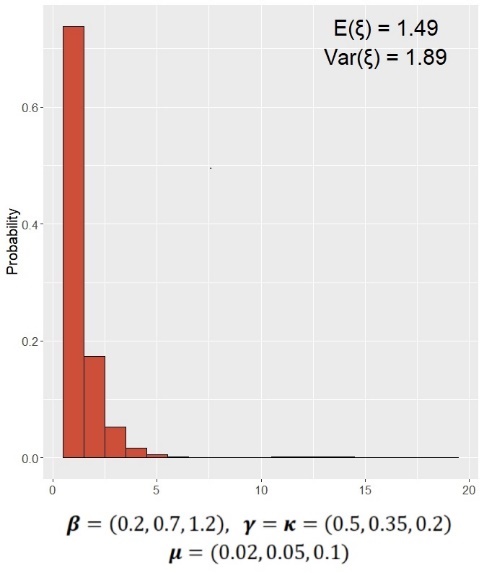}\par\medskip
\includegraphics[width=0.32\linewidth]{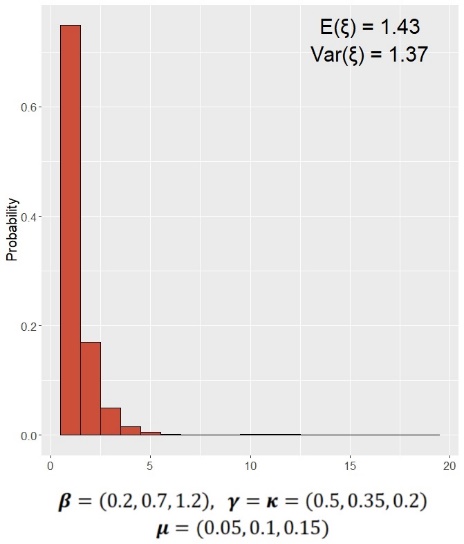}\hfill
\includegraphics[width=0.32\linewidth]{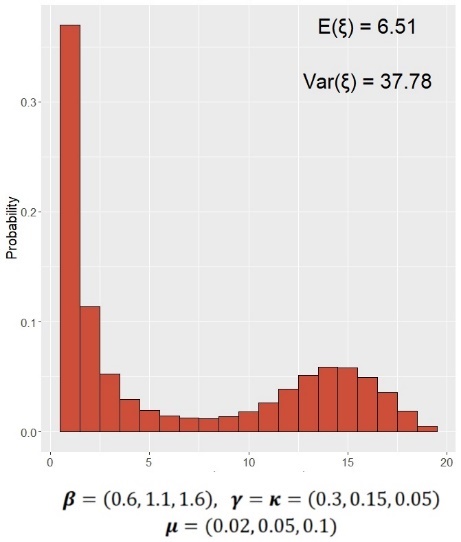}\hfill
\includegraphics[width=0.32\linewidth]{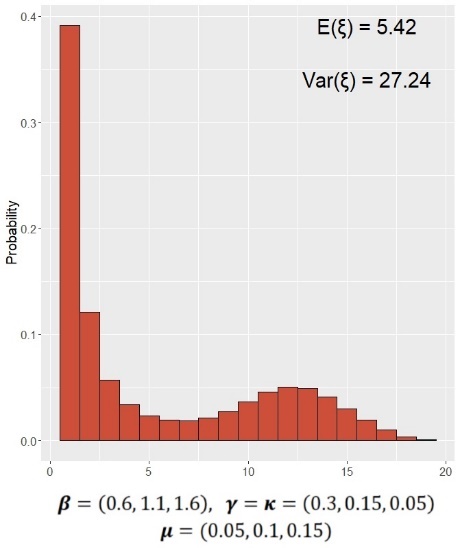}\par\medskip
\includegraphics[width=0.32\linewidth]{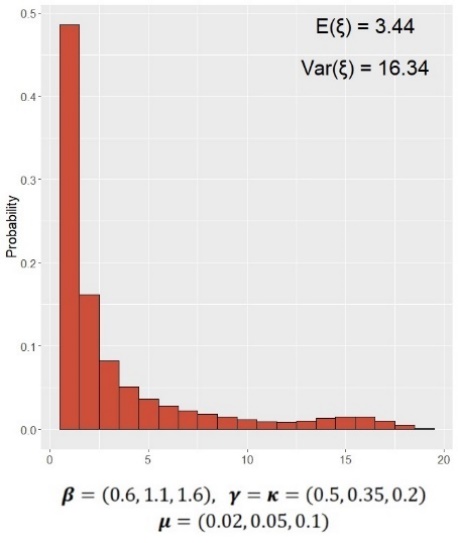}
\includegraphics[width=0.32\linewidth]{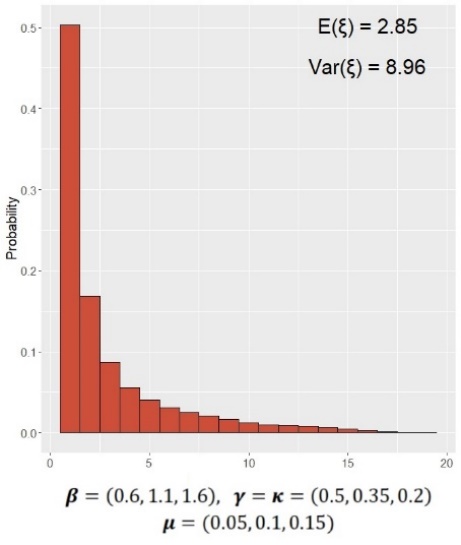}
\caption{Probability distributions of the maximum number of
simultaneously infectious cases \((\xi)\) along with the respective
expected value and variance}\label{fig:2}
\end{figure}

To validate the numerical implementation of the proposed
matrix-recursive formulas, we also considered two additional
small-population scenarios and compared the exact values obtained from
our method with direct Gillespie simulations of the CTMC. In both
scenarios, we set \(N\  = \ 20\), \(F\  = \ 3\), and the initial state
equal to \((I(0),\ A(0),\ R(0),\ F(0))\  = \ (1,\ 1,\ 5,\ 1)\). In the
first case, we use \(\bm{\beta}\  = \ (1.0,\ 1.5,\ 0.5)\),
\(p\  = \ 0.4\), \(\bm{\kappa}\  = \ (0.3,\ 0.3,\ 0.3)\),
\(\bm{\gamma}\  = \ (0.2,\ 0.2,\ 0.2)\),
\(\bm{\mu}\  = \ (0.15,\ 0.05,\ 0.01)\), with EFS transition rates
\(u_{F}(1,2) = \ u_{F}(1,3) = \frac{1}{10}\),
\(u_{F}(2,1) = \ u_{F}(2,3) = \frac{1}{15}\),
\(u_{F}(3,1) = \ u_{F}(3,2) = \frac{1}{25}\) and \(u_{F}(f,f') = \ 0\),
otherwise. In the second case, we use
\(\bm{\beta}\  = \ (1.5,\ 2.0,\ 0.8)\), \(p\  = \ 0.6\),
\(\bm{\kappa}\  = \ (0.25,\ 0.30,\ 0.35)\),
\(\bm{\gamma}\  = \ (0.15,\ 0.20,\ 0.25\)),
\(\bm{\mu}\  = \ (0.10,\ 0.05,\ 0.02)\), with EFS transition rates
\(u_{F}(1,2) = \ u_{F}(1,3) = \frac{1}{5}\),
\(u_{F}(2,1) = \ u_{F}(2,3) = \frac{1}{20}\), and
\(u_{F}(3,1) = \ u_{F}(3,2) = \frac{1}{30}\).

This comparison is not intended to show that the matrix-recursive method is merely an efficient approximation. On the contrary, the formulas developed in Sections 2.2 and 2.3 provide exact computations of the joint infection-count distribution, the maximum infectious-burden distribution, and their corresponding moments. 
The omitted probability mass is less than \(10^{-4}\) in both cases. The Gillespie algorithm,
by contrast, provides simulation-based estimates of the same quantities. Therefore, the agreement between the two approaches should be interpreted as a validation of the computational implementation of the proposed exact formulas. According to Table 1, in the two cases considered, the exact and Gillespie-based values were very close for the expected symptomatic and asymptomatic counts, as well as for the expected maximum infectious burden and standard deviations, supporting the numerical correctness of the implementation.

\Needspace{12\baselineskip}
\begin{longtable}[]{@{}
  >{\centering\arraybackslash}p{(\columnwidth - 8\tabcolsep) * \real{0.0838}}
  >{\raggedright\arraybackslash}p{(\columnwidth - 8\tabcolsep) * \real{0.2191}}
  >{\raggedright\arraybackslash}p{(\columnwidth - 8\tabcolsep) * \real{0.2100}}
  >{\centering\arraybackslash}p{(\columnwidth - 8\tabcolsep) * \real{0.2328}}
  >{\centering\arraybackslash}p{(\columnwidth - 8\tabcolsep) * \real{0.2544}}@{}}
\caption{Validation of the exact computations against Gillespie
simulations.}\label{tab:1}\\
\toprule\noalign{}
\begin{minipage}[b]{\linewidth}\raggedright
\textbf{Case}
\end{minipage} & \begin{minipage}[b]{\linewidth}\raggedright
\textbf{Quantity}
\end{minipage} & \begin{minipage}[b]{\linewidth}\raggedright
\textbf{Exact computation}
\end{minipage} & \begin{minipage}[b]{\linewidth}\raggedright
\textbf{Gillespie estimate}
\end{minipage} & \begin{minipage}[b]{\linewidth}\raggedright
\textbf{Absolute difference}
\end{minipage} \\
\midrule\noalign{}
\endhead
\bottomrule\noalign{}
\endlastfoot
1 & \(E(N^{S})\) & 16.202 & 16.156 & 0.046 \\
1 & \(E(N^{A})\) & 24.302 & 24.311 & 0.009 \\
1 & \(SD(N^{S})\) & 13.751 & 13.641 & 0.110 \\
1 & \(SD\left( N^{A} \right)\) & 21.117 & 21.138 & 0.021 \\
1 & \(E(\xi)\) & 8.833 & 8.865 & 0.032 \\
1 & \(SD(\xi)\) & 4.132 & 4.159 & 0.027 \\\midrule
2 & \(E(N^{S})\) & 35.996 & 35.979 & 0.017 \\
2 & \(E(N^{A})\) & 23.997 & 24.039 & 0.042 \\
2 & \(SD(N^{S})\) & 20.527 & 20.263 & 0.264 \\
2 & \(SD\left( N^{A} \right)\) & 14.610 & 14.453 & 0.157 \\
2 & \(E(\xi)\) & 11.341 & 11.332 & 0.009 \\
2 & \(SD(\xi)\) & 3.423 & 3.421 & 0.002 \\
\end{longtable}

\subsection{Exploring the influence of random external factors}

After conducting the sensitivity analysis outlined above, which
illustrates how the parametric space influences disease severity, we
proceed to examine the impact of transition rates between the defined
EFSs, using the model variant presented in Section 2.1.1. We present
three distinct scenarios, each representing a unique transition pattern,
with the aim of uncovering the underlying dynamics of the stochastic
system. It is important to note that, for each scenario, the remaining
parameters have been carefully selected to enable a focused examination
of the intended patterns.

\emph{Scenario A}

First, we select \((i,a,r,f) = (3,\ 3,\ 1,\ 1)\) as initial state which
is associated with a total of six currently infectious individuals. For
the three cases of this scenario, we select:

A) \(N = 30\), \(\mu = 0.05\), \(p = 0.7\), \(\beta = 0.5\),
\(\kappa = 0.5\), \(\gamma = 0.3\) and
\(u_{F}\left( f,f^{'} \right) = 0\), for \(f,f^{'} \in \{ 1, 2, 3\}\),

B) \(N = 30\), \(\mu = 0.05\),\(\ p = 0.7\),
\(\bm{\beta} = (0.5,\ 1,\ 2)\),
\(\bm{\kappa} = (0.5,\ 0.2,\ 0.1)\),
\(\bm{\gamma} = (0.3,\ 0.1,\ 0.05)\), while
\(u_{F}(1,2) = \frac{1}{15}\), \(u_{F}(2,3) = \frac{1}{5}\) and
\(u_{F}(f,f') = 0\), otherwise,

C) \(N = 30,\) \(\mu = 0.05\),\(\ p = 0.7\),
\(\bm{\beta} = (0.5,\ 1,\ 2)\),
\(\bm{\kappa} = (0.5,\ 0.2,\ 0.1)\),
\(\bm{\gamma} = (0.3,\ 0.1,\ 0.05)\), while
\(u_{F}(1,2) = \frac{1}{5}\), \(u_{F}(2,3) = \frac{1}{15}\) and
\(u_{F}(f,f') = 0\), otherwise.

The first case represents a straightforward scenario in which the
corresponding CTMC can be modeled using Equation (2), as there are no
transition rates between the EFSs. In cases (B) and (C), we deliberately
use the same parameter combinations, which are associated with
increasing disease severity. This is reflected in the rising infection
trends and the contrasting behavior of recovery rates. It is also worth
noting that the stochastic process cannot exit the third EFS. The main
distinction between these two cases is that, based on the chosen
transition rates, the process remains in the first (less severe) EFS for
a longer period in case (B).

Figures 3 and 4 present the joint probability distribution of
symptomatic/asymptomatic infections (with omitted probability mass  less than \(10^{-5}\) in all three cases), as well as the probability
distribution of the maximum number of infectious individuals, along with
the corresponding expected values and variances or standard deviations.
According to Figure 3, case (C) yields expected values of 55.47
symptomatic and 23.77 asymptomatic infections, displaying a bimodal
distribution with a more pronounced second peak compared to the milder
case (B). The higher expected number of symptomatic infections, relative
to asymptomatic cases, is consistent with the chosen symptomatic ratio
\(p = 0.7\). In case (A), which represents the least severe epidemic
scenario, the second peak of the distribution disappears, as the
infection rate remains consistently low throughout the duration of the
outbreak.

This behavior is also evident in Figure 4, where the highest expected
value corresponds to case (C). As for the joint distribution, in case
(A), the bimodal distributions observed in cases (B) and (C) are
replaced by a unimodal distribution with a peak at six infectious
individuals. It should be noted that, given the chosen values
\((i,a,r,f)\  = \ (3,\ 3,\ 1,\ 1)\), the corresponding initial state for
the modified process described in Section 2.3 is
\((i + a,a,r,f)\  = \ (6,\ 3,\ 1,\ 1)\). As a result, value six
represents the minimum possible value of this stochastic descriptor.
Finally, the highest uncertainty is observed in case (B), with
\(Var(\xi) = 61.74\). Based on the shape of the distribution,
significant probability mass is concentrated at both low and high values
of infectious individuals. This indicates the possibility of both early
epidemic eradication and trajectories where the number of infectious
individuals approaches the maximum capacity.

\begin{figure}[p]
\centering
\includegraphics[width=0.85\linewidth]{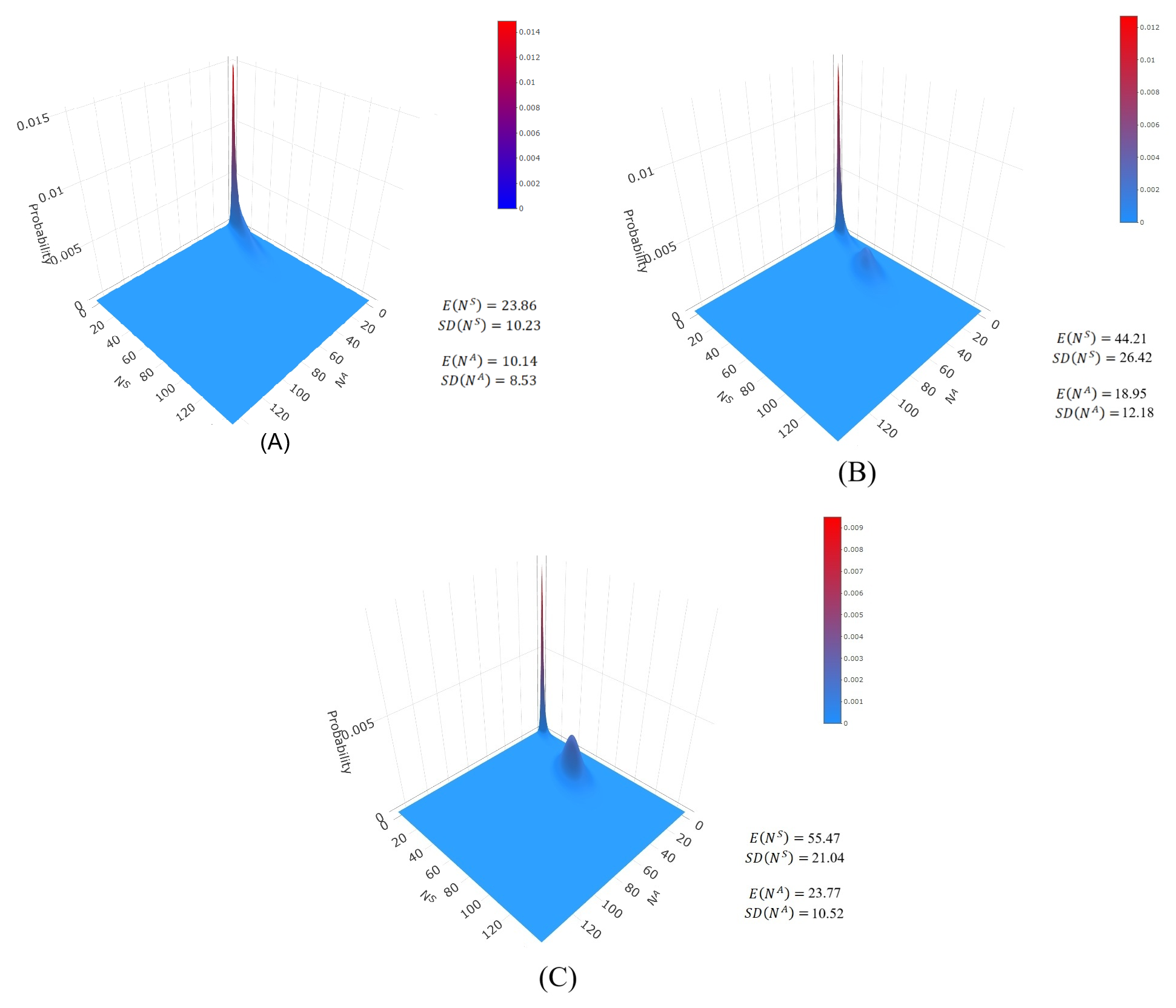}
\caption{Joint probability distributions of the total number
of new symptomatic \(\left( N^{S} \right)\) and asymptomatic \((N^{A})\)
infections for the three cases of scenario A. In each graph we include
the expected value and standard deviation of the marginal distributions
of symptomatic and asymptomatic infections}\label{fig:3}
\end{figure}

\begin{figure}[p]
\centering
\includegraphics[width=0.45\linewidth]{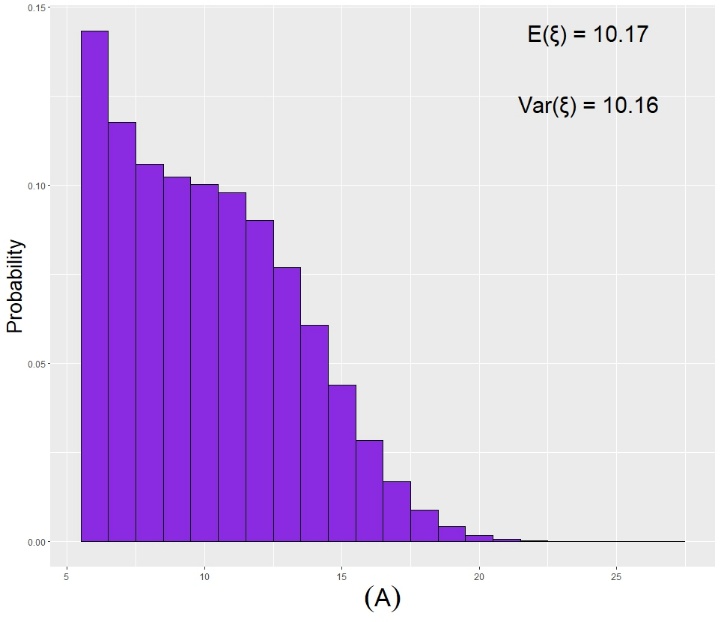}
\includegraphics[width=0.45\linewidth]{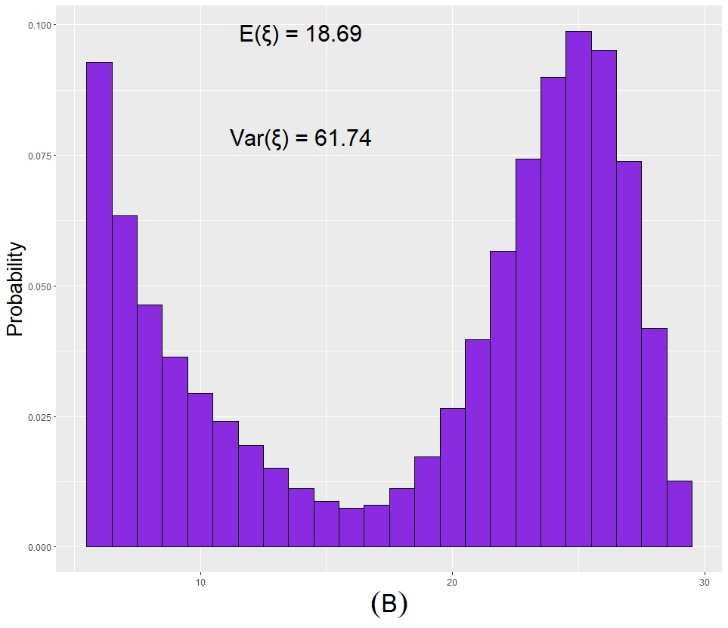}\par\medskip
\includegraphics[width=0.45\linewidth]{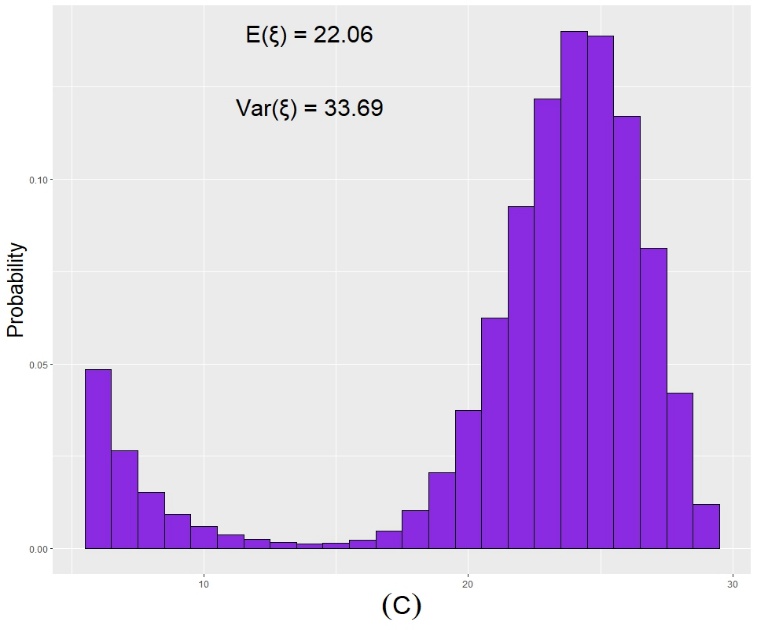}
\caption{Probability distributions of the maximum number of
simultaneously infectious cases \((\xi)\) along with the respective
expected value and variance corresponding to scenario A}\label{fig:4}
\end{figure}

\emph{Scenario B}

In this scenario, we examine the impact of the transition rates among
the EFSs while maintaining a constant population size (\(N\)) and a
fixed symptomatic ratio (\(p\)). The transmission rate decreases when
transitioning from EFS 1 to EFS 2 and then increases again when moving
from EFS 2 to EFS 3, though it remains lower than the initial rate. This
behavior may result from the implementation of a public health measure,
such as a lockdown, where the infectious force rises again after the
measure is lifted but does not return to its original level. This may be
due to partial resistance developed within the population during the
intervention period. Another contributing factor could be the
vaccination of the susceptible population with an imperfect vaccine.
While the vaccine initially reduces infectiousness, a period of
increased transmission may follow once new vaccinations stop and herd
immunity weakens due to limited protection against the virus. Finally,
it is also instructive to compare this scenario with the standard case
characterized by constant epidemic parameters.

Aiming to satisfy the aforementioned characteristics, we consider the
following three cases:

A) \(N = 30\), \(p = 0.5\), \(\beta = 2\), \(\kappa = 0.3\),
\(\gamma = 0.2\), \(\mu = 0.25\) and
\(u_{F}( f,f^{'} ) = 0\), for
\(f \neq f^{'} \in \left\{ 1,\ 2,\ 3 \right\}\),

B) \(N = 30\), \(p = 0.5\), \(\bm{\beta} = (2,\ 0.5,\ 1)\),
\(\bm{\gamma} = (0.2,\ 0.3,\ 0.4)\),
\(\bm{\kappa} = (0.3,\ 0.4,\ 0.5)\),
\(\bm{\mu} = (0.25,\ 0.15,\ 0.05)\), while
\(u_{F}(1,2) = \frac{1}{5}\), \(u_{F}(2,3) = \frac{1}{5}\) and
\(u_{F}(f,f') = 0\), otherwise,

C) \(N = 30\), \(p = 0.5\), \(\bm{\beta} = (2,\ 0.5,\ 1)\),
\(\bm{\gamma} = (0.2,\ 0.3,\ 0.4)\),
\(\bm{\kappa} = (0.3,\ 0.4,\ 0.5)\),
\(\bm{\mu} = (0.25,\ 0.15,\ 0.05)\), while
\(u_{F}(1,2) = \frac{1}{5}\), \(u_{F}(2,3) = \frac{1}{15}\) and
\(u_{F}(f,f') = 0\), otherwise.

The only difference between cases (B) and (C) remains in the second EFS
for a longer period before transitioning to the third and final EFS. We
select the initial state \((i,a,r,f) = (2,\ 2,\ 8,\ 1)\), comprising 2
symptomatic, 2 asymptomatic and 8 removed case, and begin our analysis
from the first EFS.
The omitted mass for the computation of the infection-count distribution is less than \(10^{-5}\) in all three cases.

Given that \(p = 0.5\), the infected cases are evenly distributed
between the symptomatic and asymptomatic compartments. As a result, the
joint distributions are nearly symmetric with respect to the diagonal
\(y = x\). Case (A) is associated with the highest total number of new
symptomatic and asymptomatic infections, with expected values of 35.21
(Figure 5). This outcome is attributed to the constant and high
infection rate characterizing the first EFS. Since no transitions
between EFSs are permitted, the process remains at the maximum level of
infectiousness throughout the entire period. As expected, case (C)
results in the lowest number of infections, as the stochastic process
stays in the second EFS (associated with a lower transmission rate) for
the longest period.

When considering the maximum number of simultaneously infectious cases,
case (A) yields the highest expected value, reflecting the persistently
high transmission rate in the first EFS. Cases (B) and (C) both reduce
the total and peak epidemic burden relative to case (A). According to
Figures 5 and 6, case (B) produces slightly higher expected new
infections and a slightly higher expected peak than case (C). This is
consistent with the shorter stay in the low-transmission second EFS,
which allows the process to reach the third EFS earlier.

It is also worth noting that, as in the previous scenario, all three
diagrams originate from an initial state with 4 infectious cases. The
initial state for the modified Markov chain is set to
\((i + a,a,r,f)\  = (4,2,8,1)\), meaning that the value 4 represents the
minimum value for the examined stochastic descriptor.

\begin{figure}[p]
\centering
\includegraphics[width=0.49\linewidth]{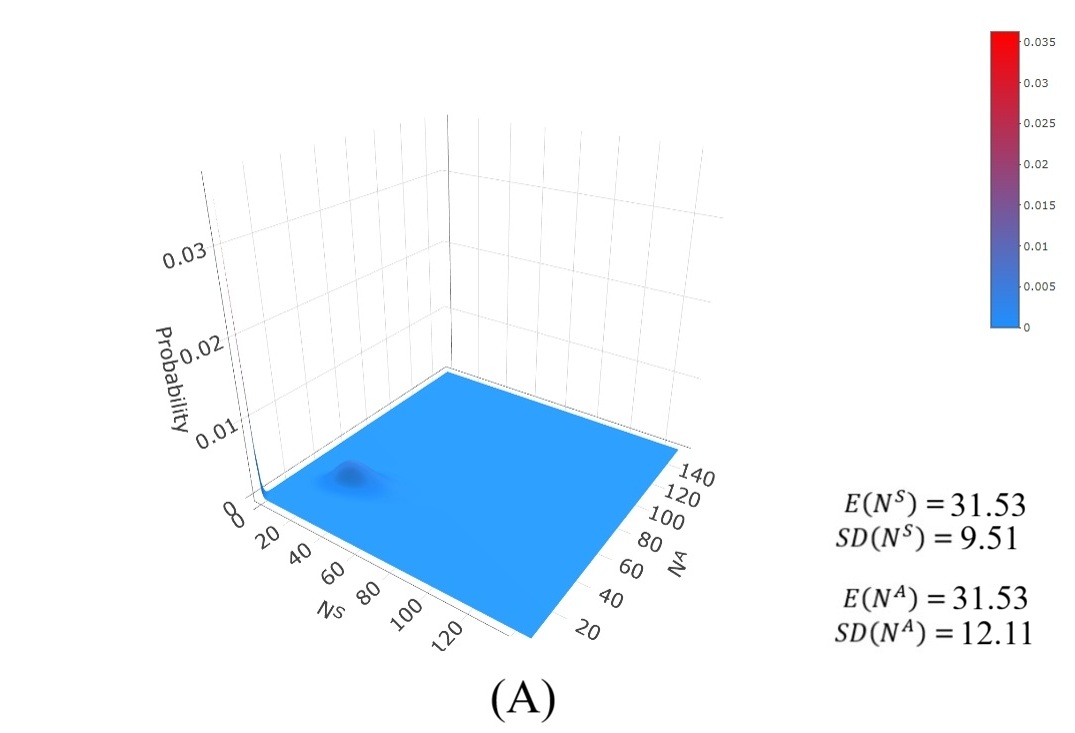}
\includegraphics[width=0.49\linewidth]{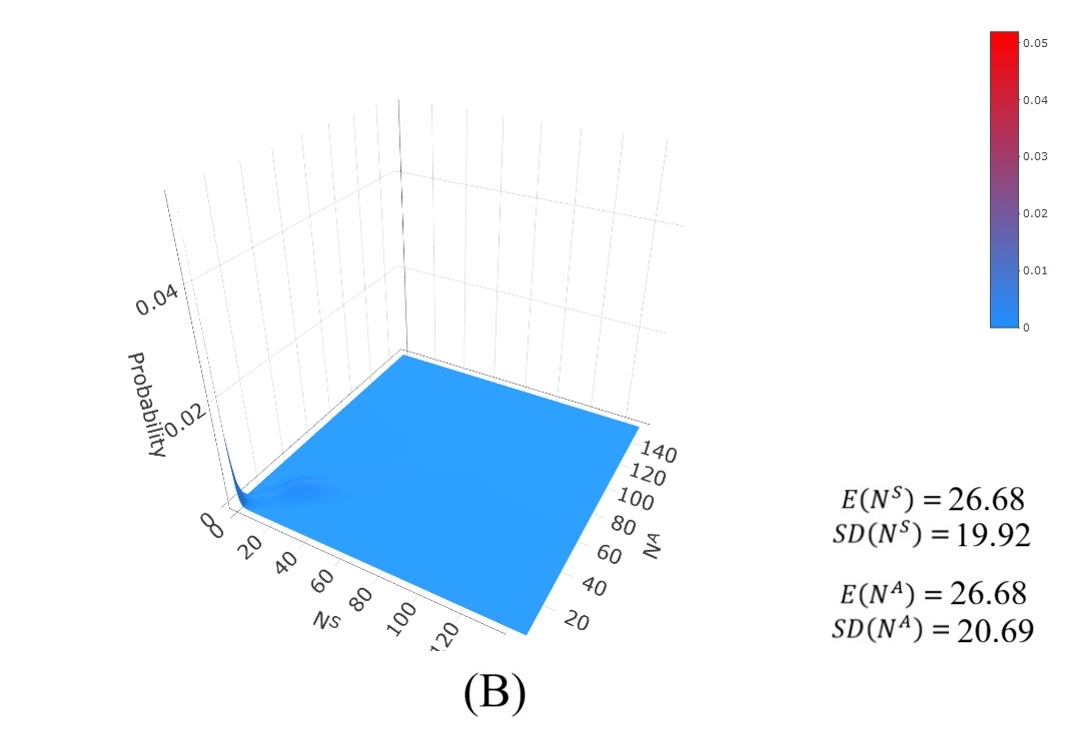}\par\medskip
\includegraphics[width=0.49\linewidth]{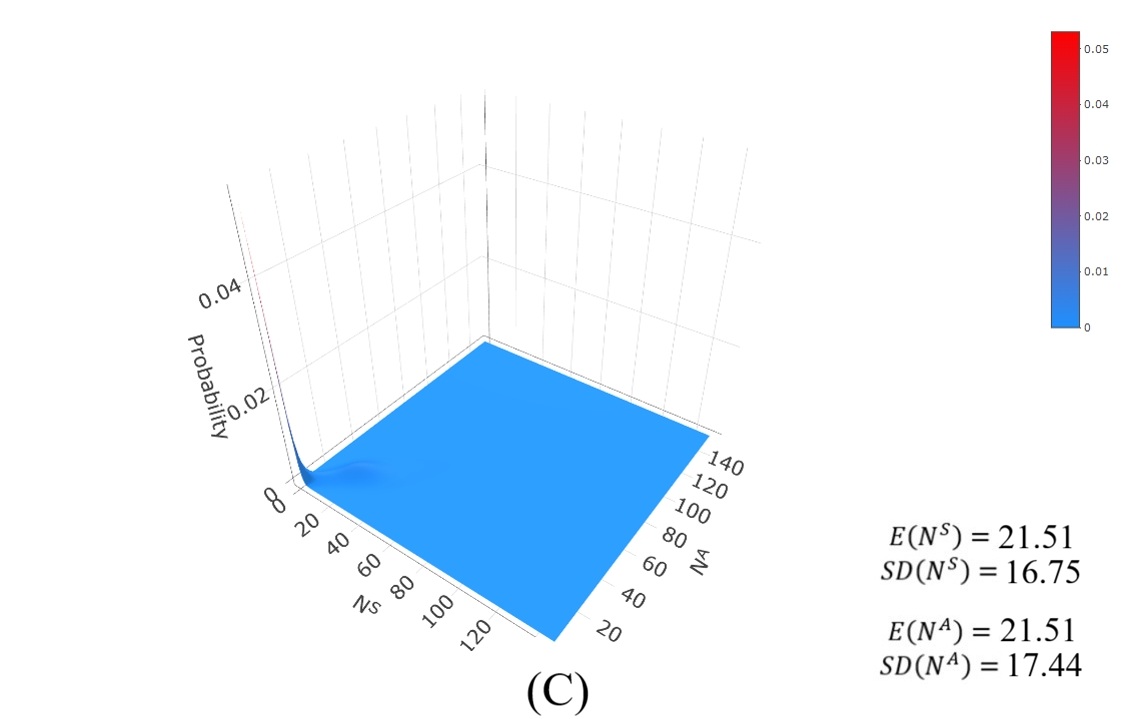}
\caption{Joint probability distributions of the total number
of new symptomatic \(\left( N^{S} \right)\) and asymptomatic \((N^{A})\)
infections for the three cases of scenario B. In each graph we include
the expected value and standard deviation of the marginal distributions
of symptomatic and asymptomatic infections}\label{fig:5}
\end{figure}

\begin{figure}[p]
\centering
\includegraphics[width=0.475\linewidth]{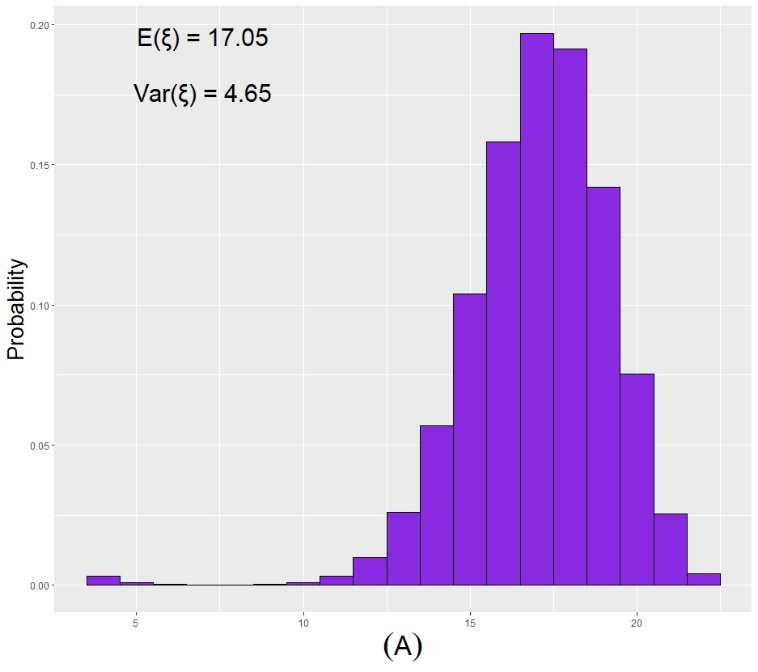}
\includegraphics[width=0.475\linewidth]{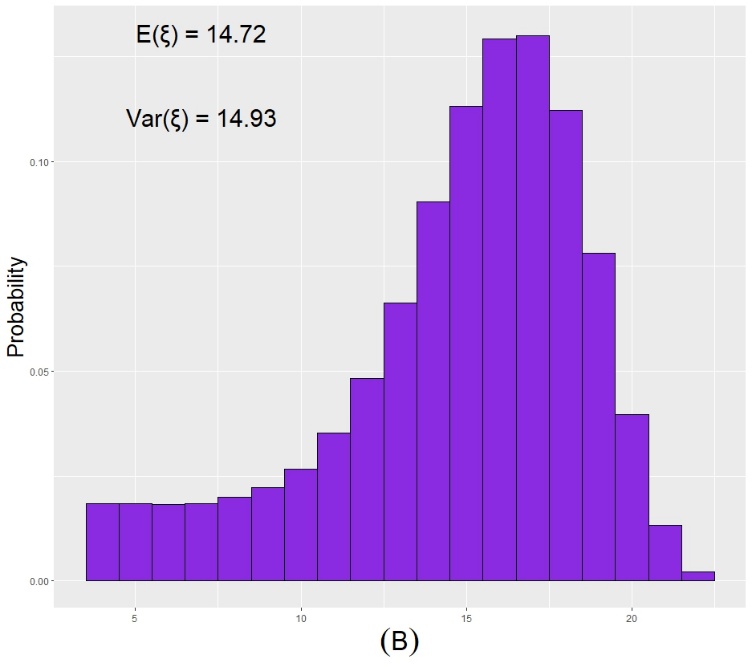}\par\medskip
\includegraphics[width=0.475\linewidth]{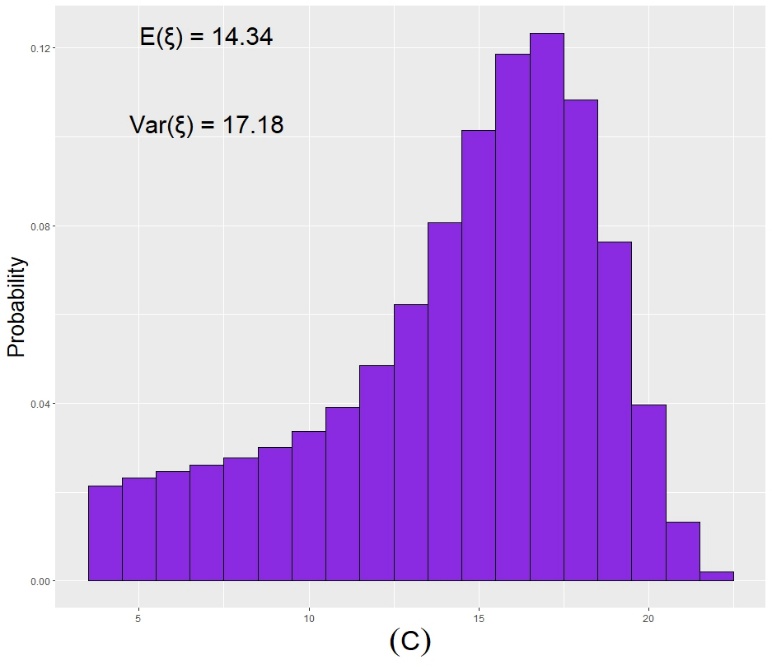}
\caption{Probability distributions of the maximum number of
simultaneously infectious cases \(\ (\xi)\) along with the respective
expected value and variance corresponding to scenario B}\label{fig:6}
\end{figure}

\emph{Scenario C}

In all cases of scenario C, we present a cyclical pattern, while we
variate the expected durations of the process to each EFS. The same
initial state as in Scenario A is selected here, too. We deem that the
external factors impact only the transmission and mortality rates. More
specifically, we set:

A) \(N = 20\), \(p = 0.4\), \(\gamma = 0.2\), \(\kappa = 0.3\),
\(\bm{\beta} = (1,\ 2,\ 0.5)\),
\(\bm{\mu} = (0.2,\ 0.15,\ 0.1)\), while
\(u_{F}(1,2) = u_{F}(1,3) = \frac{1}{5}\),
\(u_{F}(2,3) = u_{F}(2,1) = \frac{1}{15}\),
\(u_{F}(3,1) = u_{F}(3,2) = \frac{1}{25}\) and \(u_{F}(f,f') = 0\),
otherwise,

B) \(N = 20\), \(p = 0.4\), \(\gamma = 0.2\), \(\kappa = 0.3\),
\(\bm{\beta} = (1,\ 2,\ 0.5)\),
\(\bm{\mu} = (0.2,\ 0.15,\ 0.1)\), while
\(u_{F}(1,2) = u_{F}(1,3) = \frac{1}{25}\),
\(u_{F}(2,3) = u_{F}(2,1) = \frac{1}{5}\),
\(u_{F}(3,1) = u_{F}(3,2) = \frac{1}{15}\) and \(u_{F}(f,f') = 0\),
otherwise,

C) \(N = 20\), \(p = 0.4\), \(\gamma = 0.2\), \(\kappa = 0.3\),
\(\bm{\beta} = (1,\ 2,\ 0.5)\),
\(\bm{\mu} = (0.2,\ 0.15,\ 0.1)\), while
\(u_{F}(1,2) = u_{F}(1,3) = \frac{1}{15}\),
\(u_{F}(2,3) = u_{F}(2,1) = \frac{1}{25}\),
\(u_{F}(3,1) = u_{F}(3,2) = \frac{1}{5}\) and \(u_{F}(f,f') = 0\),
otherwise.

It should be noted that, in all three cases, a higher number of asymptomatic infections is observed, which is consistent with the
selected symptomatic ratio. The omitted probability mass in the cases of this scenario is less than \(10^{-7}\).

In Scenario C, the transition structure is not directed as in Scenarios
A and B. Since each EFS may have more than one outgoing transition, the
expected holding time in each EFS is computed as the reciprocal of the
total exit rate from that state, rather than from a single transition
rate.

According to Figures 7 and 8, case (B) represents the mildest instance
of an epidemic when considering both descriptors, with 19.26 symptomatic
and 28.88 asymptomatic individuals, and an expected maximum of 13.01
infectious individuals. The highest expected total number of infections
until disease extinction occurs in case (C), with 59.92 infections. This
case reflects increased epidemic severity, as based on the selected
expected durations for the EFSs, the process spends more time in the
first two, and especially in the second EFS, where the infection rate is
two to four times higher than in the other EFSs. In contrast, in case
(B), the process spends the least amount of time in the second EFS,
resulting in a reduced number of total infections.

The case associated with the highest total number of infections does not
coincide with the one exhibiting the highest infection peak. This
observation underscores the importance of analyzing both measures to
gain a more comprehensive understanding of epidemic dynamics. Finally,
the joint distributions of all three cases exhibit bimodal forms, with
the highest variance observed in case (A). This can be explained by the
shape of the distribution, where significant probability mass is
concentrated at both zero and higher numbers of total infections. This
outcome corresponds to epidemic trajectories that may either die out
quickly or persist for a considerable period.

\begin{figure}[p]
\centering
\includegraphics[width=0.45\linewidth]{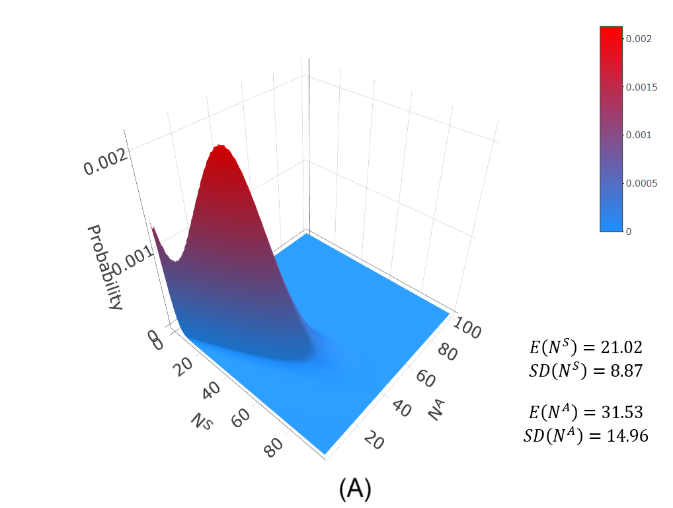}\hfill
\includegraphics[width=0.45\linewidth]{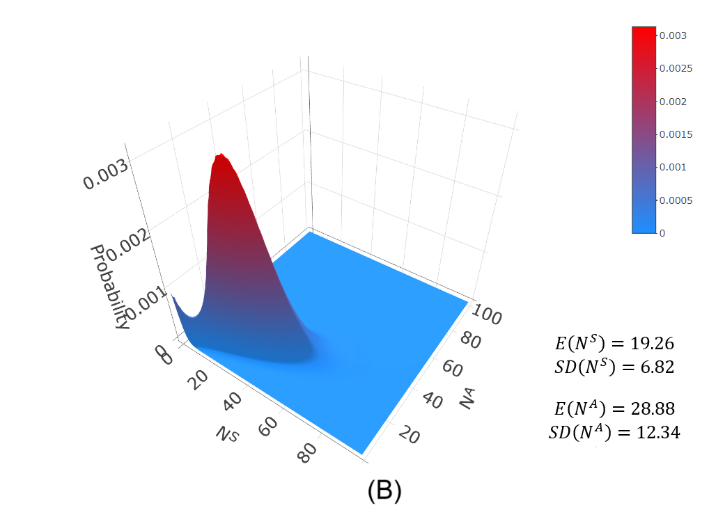}\par\medskip
\includegraphics[width=0.45\linewidth]{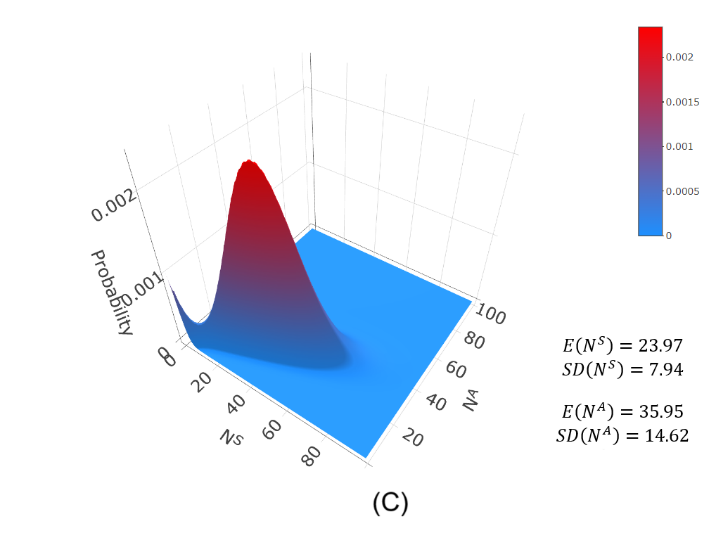}
\caption{Joint probability distributions of the total number
of new symptomatic \(\left( N^{S} \right)\) and asymptomatic \((N^{A})\)
infections for the three cases of scenario C. In each graph we include
the expected value and standard deviation of the marginal distributions
of symptomatic and asymptomatic infections}\label{fig:7}
\end{figure}

\begin{figure}[p]
\centering
\includegraphics[width=0.45\linewidth]{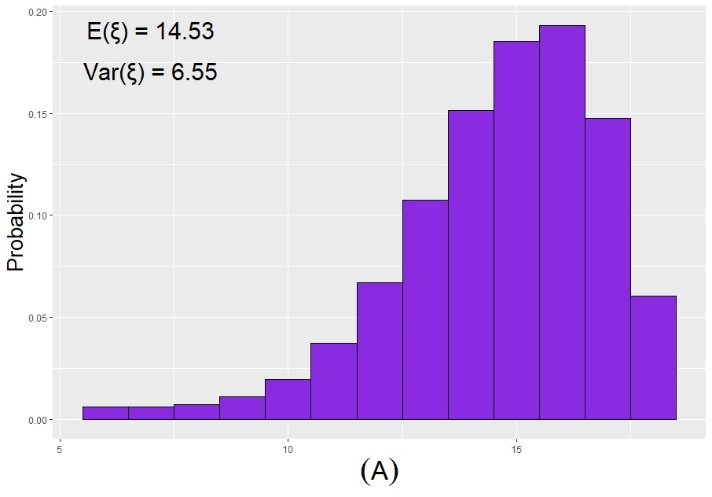}\hfill
\includegraphics[width=0.45\linewidth]{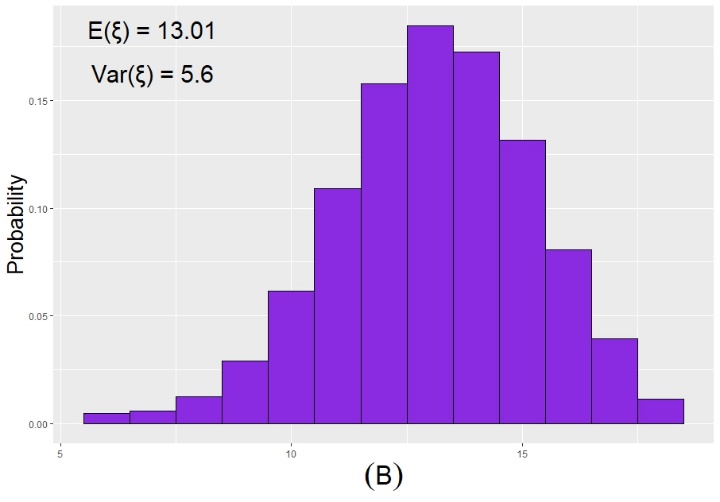}\par\medskip
\includegraphics[width=0.45\linewidth]{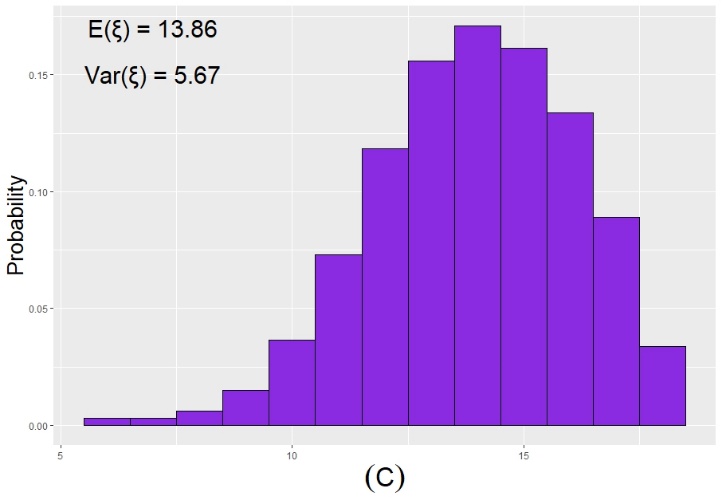}
\caption{Probability distributions of the maximum number of
simultaneously infectious cases \(\ (\xi)\) along with the respective
expected value and variance corresponding to scenario C}\label{fig:8}
\end{figure}

\FloatBarrier
\subsection{Application to 2009 H1N1 influenza pandemic}

H1N1 influenza pandemic of 2009 or Swine-Flu began its spread from
Mexico and its transmission is believed to follow a pattern similar to
that of seasonal flu \cite{ref52}. Inhalation of large infectious droplets,
along with direct contact with secretions or aerosols, facilitates
human-to-human transmission. Assessing the pandemic potential of H1N1
proves challenging due to limited data; however, epidemiological
analyses indicate that its transmission is significantly higher than
that of seasonal flu and comparable to past influenza pandemics
\cite{ref53}. The symptoms most commonly documented include fever, cough,
and sore throat. Additionally, patients mention experiencing diarrhea,
while others report vomiting, while myalgia and joint pains may also be
observed.

The majority of deaths are documented in the Americas, while Asia,
Europe and Africa have reported fewer fatalities. The primary cause of
most deaths is respiratory failure resulting from severe pneumonia and
acute respiratory distress syndrome. In Mexico, a significant portion of
deaths occurred in previously healthy individuals aged between 20 and 59
\cite{ref54}. In other parts of the world, fatalities were predominantly
observed in individuals with pre-existing health issues \cite{ref55}.

The H1N1 influenza pandemic exhibited a spatiotemporal pattern,
unfolding in three distinct waves during the spring, summer, and fall of
2009, with notable regional clusters. The initial wave, occurring in
April--May, primarily affected the greater Mexico City area and central
states. The second wave, in June 2009, was limited to the southern
states. The third wave began in August, coinciding with students
returning from summer break, and persisted through December across
Mexico \cite{ref56}.

The respective reproduction numbers were estimated as 1.8--2.1, 1.6--1.9
and 1.2--1.3, during the spring, summer and fall of 2009. According to
an analysis on influenza-like illness patients in Mexico, the 95\%
confidence interval for the case fatality ratio (CFR) for the entire
pandemic period was estimated to be 1.1--1.2\% \cite{ref56}. Influenza virus
shedding (the time during which a person might be infectious to another
person) begins the day before illness onset and can persist for 5 to 7
days, although some people may shed virus for longer periods \cite{ref57}. A
study conducted in the village of La Gloria reported that 86\% of cases
were symptomatic and ascertained, with a 95\% confidence interval of
69--100\% \cite{ref53}. It should be noted that the estimation of the
asymptomatic ratio according to data derived from small (isolated)
regions is not new \cite{ref58}, while the reported ratio necessitates the
inclusion of asymptomatic cases in the model. Finally, since the first
wave lasted from April to May, the second from June to July, and the
third from August to December, we select \(u_F(1,2) = 1/60\),
\(u_F(2,3) = 1/60\) and \(u_F( f,f^{'} ) = 0\), otherwise.

The mapping of these empirical quantities to the model parameters should
be interpreted with caution. The reported reproduction-number estimates
correspond to different regional epidemic waves in a spatially
heterogeneous setting. In the present finite-population scenario
analysis, we use them as EFS-specific values of \(R_{0}\) for a single
homogeneous closed population, which is an additional modeling
assumption. Moreover, the reported value of 86\% refers to the
proportion of cases that were symptomatic and ascertained. This quantity
is not automatically identical to the parameter \(p\), which denotes the
probability that a new infection is symptomatic. Therefore, in the
present illustration, this value is used as a literature motivated proxy
for \(p\), and the resulting numerical values should be interpreted as
scenario outputs rather than fitted estimates.

The main objective is the disease's severity with respect to the worst,
median and best-case scenarios dictated by the estimated parameter
ranges. In this application we use the immunity-preserving Variant II of
Section 2.1.2, since reinfections are not assumed to occur within the
pandemic horizon considered. The analysis should be interpreted as a
finite-population H1N1 scenario study for a small closed or semi-closed
subpopulation, where individual interactions and stochastic effects
remain relevant. The aim is to illustrate how the proposed framework
quantifies expected additional infections and peak infectious burden
under parameter values motivated by the H1N1 pandemic. For substantially
larger populations, the exact matrix-based computation becomes more
demanding because of the rapid growth of state space. In such cases,
Gillespie-type or Monte Carlo simulations may be used.

According to the above, we set \(N = 50\) and consider time-varying
transmission rate. Except for the absolute values, emphasis should be
placed on the proportion of total and maximum number of infections with
respect to the population size. According to the above, we set:

A) Best Case: \(\bm{R}_{0} = (1.8,\ 1.6,\ 1.2)\), \(CFR = 0.011\),
while \(\gamma = \kappa = \frac{1}{6}\) and \(p = 1\),

B) Median Case: \(\bm{R}_{0} = (1.95,\ 1.75,\ 1.25)\),
\(CFR = 0.0115\), while \(\gamma = \kappa = \frac{1}{6}\) and
\(p = 0.86\),

C) Worst Case: \(\bm{R}_{0} = (2.1,\ 1.9,\ 1.3)\), \(CFR = 0.012\),
while \(\gamma = \kappa = \frac{1}{6}\) and \(p = 0.69\).

Based on the case fatality ratio (CFR), where
\(CFR = \mu/(\gamma + \mu)\), we can determine the respective mortality
rate \(\mu\). So, for the three scenarios we obtain \(\mu = 0.0018\),
\(\mu = 0.0019\) and \(\mu = 0.002\). Using the reproduction numbers for
the three infection waves and considering Equation (42) of Appendix B,
we can calculate the respective transmission rates\(.\) As a result,
scenarios A, B and C, are associated with
\(\bm{\beta} = (0.303,\ 0.270,\ 0.202)\),
\(\bm{\beta} = (0.329,\ 0.295,\ 0.211)\) and
\(\bm{\beta} = (0.353,\ 0.319,\ 0.219)\), correspondingly.

Since during the examined period there is no evidence of reinfections of
previously infected individuals, we consider that patients receive
infection-induced immunity for the remaining of the pandemic. Thus, we
employ the modeling framework presented in Section 2.1.2. It should be
underlined that the expected number of symptomatic/asymptomatic
infections does not include the initial number of infectious cases.

For both descriptors we examine several initial states aiming to address
their influence on the evolution of the pandemic (Tables 2 and 3). It
should be noted that in these two tables we examine the same initial
states. For instance, state \((i,a,r,f) = (1,1,0,1)\) corresponds to the
initial state \((i + a,a,r,f) = (2,1,0,1)\) for the exploration of the
maximum number of simultaneously infectious cases.

It is evident that in the worst-case scenario it is expected to be a
higher proportion of total infections. For instance, for
\((i,a,r,f) = (1,1,0,1)\) we expect about 18.733 symptomatic and 8.416
asymptomatic cases (about 27.149 additional infections) which
corresponds to more than 50\% of the population. On the other hand, for
the best-case scenario and for the same initial state, we observe an
expected value of 21.386 symptomatic infections. We also notice that in
this case, there are no asymptomatic infections since the symptomatic
ratio (\(p)\) equals unity. The expected maximum number of
simultaneously infectious cases is 8.831 (17.7\%) and 10.997 (22.0\%)
for the best and worst-case scenario, correspondingly. The median case
presents an intermediate between the aforementioned scenarios, where for
\((i,a,r,f) = (1,1,0,1)\) we expect about 24 additional infections
(48\%) with a maximum of 9.944 (19.9\%) infectious individuals.

\Needspace{12\baselineskip}
\begin{longtable}[]{@{}
  >{\raggedright\arraybackslash}p{(\columnwidth - 12\tabcolsep) * \real{0.1197}}
  >{\raggedright\arraybackslash}p{(\columnwidth - 12\tabcolsep) * \real{0.1198}}
  >{\centering\arraybackslash}p{(\columnwidth - 12\tabcolsep) * \real{0.1645}}
  >{\centering\arraybackslash}p{(\columnwidth - 12\tabcolsep) * \real{0.1615}}
  >{\centering\arraybackslash}p{(\columnwidth - 12\tabcolsep) * \real{0.1423}}
  >{\centering\arraybackslash}p{(\columnwidth - 12\tabcolsep) * \real{0.1498}}
  >{\centering\arraybackslash}p{(\columnwidth - 12\tabcolsep) * \real{0.1423}}@{}}
\caption{Expected total number of new symptomatic/asymptomatic
infections and standard deviations for five initial states and three
epidemic scenarios}\label{tab:2}\\
\toprule\noalign{}
\begin{minipage}[b]{\linewidth}\raggedright
\end{minipage} & \begin{minipage}[b]{\linewidth}\raggedright
\end{minipage} & \begin{minipage}[b]{\linewidth}\centering

\(\bm{(1,1,0,1)}\)

\end{minipage} & \begin{minipage}[b]{\linewidth}\centering

\(\bm{(2,1,0,1)}\)

\end{minipage} & \begin{minipage}[b]{\linewidth}\centering

\(\bm{(1,2,0,1)}\)

\end{minipage} & \begin{minipage}[b]{\linewidth}\centering

\(\bm{(2,2,0,1)}\)

\end{minipage} & \begin{minipage}[b]{\linewidth}\centering

\(\bm{(2,2,2,1)}\)

\end{minipage} \\
\midrule\noalign{}
\endhead
\bottomrule\noalign{}
\endlastfoot
\multirow{4}{=}{\textbf{Best Case}} & \(E(N^{S})\) & 21.386 & 25.156 &
25.191 & 27.175 & 24.660 \\
& \(SD(N^{S})\) & 16.001 & 14.050 & 14.037 & 12.175 & 11.744 \\
& \(E(N^{A})\) & 0 & 0 & 0 & 0 & 0 \\
& \(SD(N^{A})\) & 0 & 0 & 0 & 0 & 0 \\
\multirow{4}{=}{\textbf{Median}

\textbf{Case}} & \(E(N^{S})\) & 21.042 & 24.298 & 24.328 & 25.890 &
23.611 \\
& \(SD(N^{S})\) & 14.325 & 12.175 & 12.160 & 10.310 & 10.070 \\
& \(E(N^{A})\) & 3.425 & 3.955 & 3.960 & 4.215 & 3.844 \\
& \(SD(N^{A})\) & 2.884 & 2.693 & 2.692 & 2.521 & 2.432 \\
\multirow{4}{=}{\textbf{Worst Case}} & \(E(N^{S})\) & 18.733 & 21.302 &
21.325 & 22.441 & 20.571 \\
& \(SD(N^{S})\) & 11.807 & 9.789 & 9.775 & 8.177 & 8.062 \\
& \(E(N^{A})\) & 8.416 & 9.570 & 9.581 & 10.082 & 9.242 \\
& \(SD(N^{A})\) & 5.730 & 4.967 & 4.962 & 4.370 & 4.273 \\
\end{longtable}

\Needspace{12\baselineskip}
\begin{longtable}[]{@{}
  >{\raggedright\arraybackslash}p{(\columnwidth - 12\tabcolsep) * \real{0.1207}}
  >{\raggedright\arraybackslash}p{(\columnwidth - 12\tabcolsep) * \real{0.1388}}
  >{\centering\arraybackslash}p{(\columnwidth - 12\tabcolsep) * \real{0.1495}}
  >{\centering\arraybackslash}p{(\columnwidth - 12\tabcolsep) * \real{0.1516}}
  >{\centering\arraybackslash}p{(\columnwidth - 12\tabcolsep) * \real{0.1536}}
  >{\centering\arraybackslash}p{(\columnwidth - 12\tabcolsep) * \real{0.1429}}
  >{\centering\arraybackslash}p{(\columnwidth - 12\tabcolsep) * \real{0.1429}}@{}}
\caption{Expected maximum number of simultaneously infectious
cases, standard deviations and probabilities that the maximum number
exceeds \(c = 5\), \(c = 10\) and \(c = 20\), for five initial states
and three epidemic scenarios}\label{tab:3}\\
\toprule\noalign{}
\begin{minipage}[b]{\linewidth}\raggedright
\end{minipage} & \begin{minipage}[b]{\linewidth}\raggedright
\end{minipage} & \begin{minipage}[b]{\linewidth}\centering

\(\bm{(2,1,0,1)}\)

\end{minipage} & \begin{minipage}[b]{\linewidth}\centering

\(\bm{(3,1,0,1)}\)

\end{minipage} & \begin{minipage}[b]{\linewidth}\centering

\(\bm{(3,2,0,1)}\)

\end{minipage} & \begin{minipage}[b]{\linewidth}\centering

\(\bm{(4,2,0,1)}\)

\end{minipage} & \begin{minipage}[b]{\linewidth}\centering

\(\bm{(4,2,2,1)}\)

\end{minipage} \\
\midrule\noalign{}
\endhead
\bottomrule\noalign{}
\endlastfoot
\multirow{5}{=}{\textbf{Best Case}} & \(E(\xi)\) & 8.831 & 10.463 &
10.476 & 11.565 & 10.815 \\
& \(SD(\xi)\) & 5.451 & 4.930 & 4.930 & 4.529 & 4.281 \\
& \(P(\xi > 5)\) & 0.645 & 0.791 & 0.792 & 0.887 & 0.869 \\
& \(P(\xi > 10)\) & 0.401 & 0.502 & 0.503 & 0.583 & 0.515 \\
& \(P(\xi > 20)\) & 0.016 & 0.022 & 0.022 & 0.028 & 0.016 \\
\multirow{5}{=}{\textbf{Median}

\textbf{Case}} & \(E(\xi)\) & 9.944 & 11.610 & 11.624 & 12.686 &
11.839 \\
& \(SD(\xi)\) & 5.865 & 5.208 & 5.207 & 4.743 & 4.506 \\
& \(P(\xi > 5)\) & 0.695 & 0.831 & 0.832 & 0.914 & 0.899 \\
& \(P(\xi > 10)\) & 0.491 & 0.598 & 0.599 & 0.676 & 0.611 \\
& \(P(\xi > 20)\) & 0.032 & 0.041 & 0.041 & 0.051 & 0.030 \\
\multirow{5}{=}{\textbf{Worst Case}} & \(E(\xi)\) & 10.997 & 12.679 &
12.694 & 13.721 & 12.793 \\
& \(SD(\xi)\) & 6.190 & 5.407 & 5.404 & 4.887 & 4.667 \\
& \(P(\xi > 5)\) & 0.732 & 0.861 & 0.862 & 0.933 & 0.921 \\
& \(P(\xi > 10)\) & 0.569 & 0.676 & 0.678 & 0.749 & 0.689 \\
& \(P(\xi > 20)\) & 0.053 & 0.067 & 0.067 & 0.081 & 0.050 \\
\end{longtable}

Regarding the initial states examined, it is evident that the state
\((i,a,r,f) = (2,2,0,1)\) is associated with the highest expected total
and peak number of infectious cases across all three scenarios. This can
be attributed to the higher number of symptomatic and asymptomatic
individuals, combined with the absence of removed individuals, allowing
all to contribute to the transmission of the virus. Consequently, the
state \((i,a,r,f) = (2,2,2,1)\) results in lower expected infection
numbers compared to \((i,a,r,f) = (2,2,0,1)\).

The low number of infectious individuals in the state
\((i,a,r,f) = (1,1,0,1)\), may lead to the premature extinction of the
disease, which explains why this state is associated with the lowest
expected number of infections. Finally, we observe that the state
\((i,a,r,f) = (1,2,0,1)\) generates slightly more infections than
\((i,a,r,f) = (2,1,0,1)\), even though both are characterized by three
infectious individuals. This pattern is consistent across all three
scenarios and for both stochastic descriptors. From this, we conclude
that the presence of asymptomatic cases is more burdensome for the
population. This is likely because asymptomatic individuals exhibit only
mild or no symptoms and typically do not progress to mortality. As a
result, they remain in the infectious state for a longer duration,
thereby contributing to sustained transmission of the disease.

\FloatBarrier
\section{Discussion}

In this paper, we introduce a stochastic SuASR framework that accounts
for the influence of random external factors. The analysis is based on
an extended CTMC that captures the effects of external factors, which
can significantly affect the parameters governing the epidemic dynamics.
The proposed stochastic framework accounts for two sources of infection,
originating from either asymptomatic or symptomatic individuals, leading
to more complex and insightful dynamics. Notably, the integration of a
Markovian model that incorporates asymptomatic infections along with the
impact of external factors constitutes a novel contribution to
literature. Recognizing the significant roles both symptomatic and
asymptomatic individuals play in disease transmission, we introduce new
computational methods for efficiently calculating their joint
distribution, which represents the driving force of the model.

By modifying the initially constructed four-dimensional chain, we derive
formulas for computing extreme values, such as the maximum number of
simultaneously infectious cases prior to disease eradication. Overall,
the maximum number along with the total number of new
symptomatic/asymptomatic infections offer a comprehensive depiction of
the epidemic's severity, capturing both the total number of infections
from each source and the occurrence of extreme cases. The proposed
framework comprises two related formulations, allowing either
recurrent-infection dynamics or immunity-preserving dynamics according
to the epidemiological setting.

The computation of the joint distribution for the total number of new
symptomatic/asymptomatic infections provides a significant advantage. It
enables the derivation of formulas for the marginal distributions of
each infection type. Specifically, we have
\(\widetilde{\bm{p}}(n) = \sum_{m \in \mathbb{N}_{0}}^{}{\widetilde{\bm{p}}(n,m)}\)
and
\(\widetilde{\bm{p}}(m) = \sum_{n \in \mathbb{N}_{0}}^{}{\widetilde{\bm{p}}(n,m)}\),
where \(\mathbb{N}_{0}\mathbb{= N \cup}\left\{ 0 \right\}\). Apparently,
the computation of infinite terms is infeasible. However, quantities
\(\sum_{m = m_{0} + 1}^{\infty}{\widetilde{\bm{p}}(n,m)}\) and
\(\sum_{n = n_{0} + 1}^{\infty}{\widetilde{\bm{p}}(n,m)}\) become
negligible for \(m_{0}\) and \(n_{0}\) sufficiently large; thus
\(\sum_{m = m_{0} + 1}^{\infty}{\widetilde{\bm{p}}(n,m)} \approx \varepsilon(m_{0})\ \)
and
\(\sum_{n = n_{0} + 1}^{\infty}{\widetilde{\bm{p}}(n,m)} \approx \varepsilon\left( n_{0} \right)\).
Hence, we obtain

\begin{equation*}
\widetilde{\bm{p}}(n) = \sum_{m \in \mathbb{N}_{0}}^{}{\widetilde{\bm{p}}(n,m)} \approx \sum_{m = 0}^{m_{0}}{\widetilde{\bm{p}}(n,m)} + \varepsilon\left( m_{0} \right) \approx \sum_{m = 0}^{m_{0}}{\widetilde{\bm{p}}(n,m)},
\end{equation*}
and
\begin{equation*}
\widetilde{\bm{p}}(m) = \sum_{n \in \mathbb{N}_{0}}^{}{\widetilde{\bm{p}}(n,m)} \approx \sum_{n = 0}^{n_{0}}{\widetilde{\bm{p}}(n,m)} + \varepsilon\left( n_{0} \right) \approx \sum_{n = 0}^{n_{0}}{\widetilde{\bm{p}}(n,m)},
\end{equation*}

that can be used to estimate the respective marginal probability
distributions.

In the Methodology section, we present alternative representations for
computing the joint distribution, each of which is more suitable
depending on the population size \((N)\). Furthermore, we recommend
solving the resulting matrix equations, using methods such as LU
factorization, without explicitly computing the inverse matrix, in order
to minimize the number of operations. Additionally, storing
high-dimensional matrices in sparse format is advisable, as they are
predominantly composed of zeros due to their block tridiagonal
structure. Adopting this approach significantly reduces the system's
memory requirements.

Regarding the Results section, an extensive sensitivity analysis
containing several illustrative examples is provided. Emphasis is placed
on both the examination of the system's behavior concerning the
epidemic's parameters or the transition rates among the EFSs and the
application of the methodology based on parameters derived from a
real-world outbreak. Higher transmission (\(\beta)\) and lower recovery
\((\gamma,\kappa)\) and mortality \((\mu)\) rates are associated with
prolonged epidemic durations, and consequently greater numbers of total
and maximum infections, as infected individuals remain for longer
periods in compartments facilitating the spread of the disease. This
observation is consistent with the results of \cite{ref12,ref24,ref26,ref59,ref60,ref61}.

An interesting observation is that it is possible for two epidemic
scenarios to differ in such a way that one results in a higher total
number of infections while exhibiting a lower peak number of infectious
individuals compared to the other. In scenarios where the spread of
infection occurs gradually, due to factors such as moderate transmission
rates or the implementation of control measures, the epidemic tends to
persist over a longer duration. As a result, although the total number
of infections may be higher, the infectious burden remains relatively
low, producing a flattened epidemic curve. Conversely, in scenarios
characterized by rapid transmission and limited intervention, the number
of infectious individuals can rise sharply to a higher peak, but the
epidemic may subside quickly due to rapid recovery or the depletion of
susceptible individuals, ultimately leading to a lower total number of
cases. These distinctions underscore the importance of considering both
the peak burden and the total disease impact when evaluating and
comparing epidemic outcomes.

The necessity of investigating epidemic dynamics through stochastic, and
not deterministic models is highlighted. In cases of bimodal
distributions, the information provided by the mean estimates given by
an ODE system is insufficient. The existence of two peaks signifies the
existence of scenarios where the epidemic may die out prematurely or
remain consistent over time. These scenarios, accompanied by high
standard deviation necessitate more attention and close inspection of
the epidemic's evolution due to significant uncertainty. Public health
authorities can use this information to adjust the implementation of
control measures and help contain the spread of the disease.

Finally, it is worth noting that certain modeling efforts have explored
transition periods with generally distributed durations using
piecewise-deterministic Markov processes \cite{ref62,ref63}. However, beyond
these approaches, there is a notable lack of analysis on the impact of
non-exponential transition times within the Markov framework. To address
this gap, we propose integrating semi-Markov processes, which allow for
more flexible sojourn time distributions and may offer valuable insights
to existing literature. Finally, emphasis could be placed on change
point analysis (CPA) applied to real epidemic data from novel disease
outbreaks. CPA enables the detection of shifts in the dynamics of the
outbreak \cite{ref64,ref65,ref66}, allowing to identify transitions among EFSs.
This, in turn, will facilitate the development of a statistical
estimation approach suitable for both offline and online prediction.

\section*{Declaration of Competing Interest}

The authors declare that they have no known competing financial
interests or personal relationships that could have appeared to
influence the work reported in this paper.

\section*{Data availability Statement}

The code used for the simulation study is publicly available at
\url{https://github.com/VasilisPapageorgiou/markovian-suasr}.

\section*{Funding}

This research was funded by the A. G. Leventis Foundation through a
doctoral studies grant.

\section*{Ethics Approval}

This study did not involve human participants, human material, or
identifiable personal data. It is based on mathematical modeling and
publicly available/aggregate data.

\section*{Appendix A}

Aiming to prove formula (19) and in order to construct an algorithm for
the computation of the \(\bm{p}(n,m)\) vector, we define the
generating functions \(\varphi_{i,a,r,f}(z,s)\) for \(|z|,|s| \leq 1\)
and \((i,a,r,f) \in A_T\), where based on first-step analysis \cite{ref67}, we
obtain the formula

\begin{equation}
\begin{aligned}
&\ \varphi_{i,a,r,f}(z,s) = z\frac{p(f)}{{\widetilde{u}}_{i,a,r,f}}\left( \frac{\beta(f)}{N} \right)_{i,a,r}\varphi_{i + 1,a,r,f}(z,s)\left( 1 - \delta_{N - i - a - r,0} \right)\ \\
&\quad + s\frac{1 - p(f)}{{\widetilde{u}}_{i,a,r,f}}\left( \frac{\beta(f)}{N} \right)_{i,a,r}\varphi_{i,a + 1,r,f}(z,s)\left( 1 - \delta_{N - i - a - r,0} \right) \\
&\quad + \frac{\mu_{i}(f)}{{\widetilde{u}}_{i,a,r,f}}\varphi_{i - 1,a,r + 1,f}(z,s)\left( 1 - \delta_{i0} \right) + \frac{\gamma_{i}(f)}{{\widetilde{u}}_{i,a,r,f}}\varphi_{i - 1,a,r,f}(z,s)\left( 1 - \delta_{i0} \right) \\
&\quad + \frac{\kappa_{a}(f)}{{\widetilde{u}}_{i,a,r,f}}\varphi_{i,a - 1,r,f}(z,s)\left( 1 - \delta_{a0} \right) + \sum_{\substack{f'=1\\f'\ne f}}^F\frac{u_{F}( f,f^{'} )}{{\widetilde{u}}_{i,a,r,f}}\varphi_{i,a,r,f'}(z,s)\ .
\end{aligned}\tag{31}\label{eq:31}
\end{equation}

After multiplying both sides with the denominator
\({\widetilde{u}}_{i,a,r,f}\) and expressing (31) in matrix form, we
obtain

\begin{equation}
\widetilde{\bm{U}}(z,s)\bm{\varphi}(z,s) = - \left( \delta_{i,1}\delta_{a,0}\left( \mu_{1}(f) + \gamma_{1}(f) \right) + \delta_{i,0}\delta_{a,1}\ \kappa_{1}(f) \right)\bm{1}_{\left| A_{T} \right|},\tag{32}\label{eq:32}
\end{equation}

considering that for \(i = 1\) and \(a = 0\),

\begin{equation}
\begin{aligned}
&zp(f)\left(\frac{\beta(f)}{N}\right)_{1,0,r}\varphi_{2,0,r,f}(z,s)
 \bigl(1-\delta_{N-1-r,0}\bigr)+\sum_{\substack{f'=1\\f'\ne f}}^F u_F(f,f')\varphi_{1,0,r,f'}(z,s)\\
&\quad+s\bigl(1-p(f)\bigr)\left(\frac{\beta(f)}{N}\right)_{1,0,r}
 \varphi_{1,1,r,f}(z,s)\bigl(1-\delta_{N-1-r,0}\bigr)\\
&\quad-\widetilde{u}_{1,0,r,f}\varphi_{1,0,r,f}(z,s)
 =-\bigl(\mu_1(f)+\gamma_1(f)\bigr),\qquad 1\le f\le F,
\end{aligned}
\tag{33}\label{eq:33}
\end{equation}

as \(\varphi_{0,0,r,f}(z,s) = 1\), due to disease's extinction. The same
applies for the case \(i = 0\) and \(a = 1\), where

\begin{equation}
\begin{aligned}
&zp(f)\left( \frac{\beta(f)}{N} \right)_{0,1,r}\varphi_{1,1,r,f}(z,s)\left( 1 - \delta_{N - 1 - r,0} \right) + \sum_{\substack{f'=1\\f'\ne f}}^F u_{F}( f,f^{'} )\varphi_{0,1,r,f^{'}}(z,s) \\
&+ s\left( 1 - p(f) \right)\left( \frac{\beta(f)}{N} \right)_{0,1,r}\varphi_{0,2,r,f}(z,s)\left( 1 - \delta_{N - 1 - r,0} \right)\  - {\widetilde{u}}_{0,1,r,f}\varphi_{0,1,r,f}(z,s) \\
&= - \kappa_{1}(f)\ ,\quad 1 \leq f \leq F.
\end{aligned}\tag{34}\label{eq:34}
\end{equation}

We note that the polynomial matrix
\(\widetilde{\bm{U}}(z,s) = \{{\widetilde{u}}_{(i,a,r,f),( i^{'},a^{'},r^{'},f^{'} )}(z,s): (i,a,r,f),( i^{'},a^{'},r^{'},f^{'} ) \in A_{T}\}\),
has the form

\begin{equation}
{\widetilde{u}}_{(i,a,r,f),( i^{'},a^{'}, r^{'},f^{'} )}(z,s) = \begin{cases}
- {\widetilde{u}}_{i,a,r,f},   & ( i^{'},a^{'}, r^{'},f^{'} ) = (i,a,r,f), \\
zp(f)\left( \frac{\beta(f)}{N} \right)_{i,a,r},   & ( i^{'},
a^{'}, r^{'},f^{'} ) = (i + 1,a,r,f), \\
s\left( 1 -  p(f) \right)\left( \frac{\beta(f)}{N} \right)_{i,a,r},   & ( i^{'},a^{'}, r^{'},f^{'} ) = (i,a + 1,r,f), \\
\gamma_{i}(f),   & ( i^{'},a^{'}, r^{'},f^{'} ) = (i - 1,a,r,f), \\
\kappa_{a}(f),  & ( i^{'},a^{'}, r^{'},f^{'} ) = (i,a - 1,r,f), \\
{  \mu}_{i}(f),   & ( i^{'},a^{'}, r^{'},f^{'} ) = (i - 1,a,r + 1,f) , \\
u_{F}\left( f,f^{'} \right),   & ( i^{'},a^{'}, r^{'},f^{'} ) = (i,a,r,f'),  f \neq f^{'}, \\
0,   & \text{otherwise}.

\end{cases}\tag{35}\label{eq:35}
\end{equation}

\begin{longtable}[]{@{}
  >{\raggedright\arraybackslash}p{(\columnwidth - 0\tabcolsep) * \real{1.0000}}@{}}
\toprule\noalign{}
\begin{minipage}[b]{\linewidth}\raggedright
\textbf{Algorithm.} Joint probability distribution of the total number
of new symptomatic/asymptomatic cases
\end{minipage} \\
\midrule\noalign{}
\endhead
\bottomrule\noalign{}
\endlastfoot
\textbf{Input:} \(N\), \(\bm{\kappa}\), \(\bm{\beta}\),
\(\bm{\mu}\), \(\bm{p}\), \(\bm{\gamma}\),
\(\bm{u}_{F}\), \(n_{\max}\), \(m_{\max}\).

\textbf{Step 1:} Compute \(\widetilde{\bm{p}}(0,0)\).

Step 1a: Construct the polynomial matrix \(\widetilde{\bm{U}}(z,s)\)
based on (35).

Step 1b: Set \(z = s = 0\) and solve

\(\widetilde{\bm{U}}(0,0)\widetilde{\bm{p}}(0,0) = - \left( \delta_{i,1}\delta_{a,0}\left( \mu_{1}(f) + \gamma_{1}(f) \right) + \delta_{i,\ 0}\delta_{a,\ 1}\kappa_{1}(f) \right)\bm{1}_{\left| A_{T} \right|}.\)

\textbf{Step 2:} Set \(m = 0.\) Compute \(\widetilde{\bm{p}}(n,0)\)
for \(1 \leq n \leq n_{\max}.\)

Step 2a: Construct matrices \({\widetilde{\bm{U}}}_{T}^{*}\) and
\({\widetilde{\bm{U}}}_{T}^{+},\ \)according to (20) and (21).

Step 2b: For \(n = 1,\ldots,\ n_{\max}\)

\({\widetilde{\bm{U}}}_{T}^{*}\widetilde{\bm{p}}(n,0) = - {\widetilde{\bm{U}}}_{T}^{+}\widetilde{\bm{p}}(n - 1,\ 0)\ .\)

\textbf{Step 3:} Set \(n = 0.\) Compute \(\widetilde{\bm{p}}(0,m)\)
for \(1 \leq m \leq m_{\max}.\)

Step 3a: Construct matrices \({\widetilde{\bm{U}}}_{T}^{*}\) and
\({\widetilde{\bm{U}}}_{T}^{-},\ \)according to (20) and (22).

Step 3b: For \(m = 1,\ldots,m_{\max}\)

\({\widetilde{\bm{U}}}_{T}^{*}\widetilde{\bm{p}}(0,m) = - {\widetilde{\bm{U}}}_{T}^{-}\widetilde{\bm{p}}(0,m - 1)\ .\)

\textbf{Step 4:} Compute the 3-dimensional tensor
\(\widetilde{\bm{p}}(n,m)\) for \(1 \leq n \leq n_{\max}\) and
\(1 \leq m \leq m_{\max}.\)

Step 4a: Construct matrices \({\widetilde{\bm{U}}}_{T}^{*}\),
\({\widetilde{\bm{U}}}_{T}^{+}\) and
\({\widetilde{\bm{U}}}_{T}^{-},\ \)according to (20)--(22).

Step 4b: For \(m = 1,\ldots,m_{\max}\) and \(n = 1,\ldots,n_{\max}\)

\({\widetilde{\bm{U}}}_{T}^{*}\widetilde{\bm{p}}(n,m) = - \left( {\widetilde{\bm{U}}}_{T}^{+}\widetilde{\bm{p}}(n - 1,m) + {\widetilde{\bm{U}}}_{T}^{-}\widetilde{\bm{p}}(n,m - 1) \right).\)

\textbf{Output}: Tensor \(\widetilde{\bm{p}}(n,m)\). \\
\end{longtable}

The column vector \(\bm{\varphi}(z,s)\) includes all the generating
functions corresponding to the set of the system's transient states. By
setting \(z = s = 0\) in (32), we generate the vector of the joint
distribution \(\widetilde{\bm{p}}(0,0).\ \)The following algorithm
presents the calculation of the 3-dimensional tensor
\(\widetilde{\bm{p}}(n,m)\), based on the recursive formula (19)
presented in the Remark. The rows and columns of the tensor correspond
to the total number of new symptomatic and asymptomatic cases,
respectively, while the third dimension is associated with the selected
current state of the system.

The above algorithm is presented for Variant I. For Variant II, the same
computational procedure applies, with the only modifications concerning
the recovery-related transition blocks, which must be defined according
to the immunity-preserving formulation of Section 2.1.2.

\section*{Appendix B}

In the experimental part of the present paper, we examine the spread of
the H1N1 influenza pandemic of 2009, using the model variant presented
in Section 2.1.2. In this section, the selection of realistic values for
the system's parameters is necessary. The transmission rate \((\beta)\)
can be retrieved from the widely utilized basic reproduction number
\({(R}_{0})\). This index represents the expected number of secondary
infections generated by an already infected individual inside a nearly
susceptible population.

We can derive a formula for this index based on the methodology of the
next-generation matrix which was developed by \cite{ref68} and was
popularized by \cite{ref69}. Following the process described in \cite{ref39,ref69},
we first define the disease-free equilibrium (DFE)
\(\bm{Y}^{0} = \left( S^{0},\ I^{0},A^{0} \right) = (N,0,0).\) This
vector is associated with the absence of the virus from the examined
population. We notice that the presented epidemiological scheme contains
a system of 4 differential equations. However, we present only 3
components for \(\bm{Y}^{0}\). This modification derives from the
fact that the number of removed cases is a linear combination of the
other 3 states, since the examined population is closed.

Next, we select the vector \(\bm{X} = (I,A)\) which contains the two
states of the system which are associated with the existence and spread
of the virus. Let \(\bm{\mathcal{F}}\left( \bm{X} \right)\) and
\(\bm{\mathcal{V}}(\bm{X})\) be 2×1 vectors whose \(k\)-th entry exhibits the rate
of new infections entering state \(k\) and the transfer rate out of
state \(k\) by non-infection means, respectively. As a result,

\begin{equation}
{\dot{\bm{X}}}^{T}=\bm{\mathcal{F}}\left( \bm{X} \right) - \bm{\mathcal{V}}\left( \bm{X} \right),\tag{36}\label{eq:36}
\end{equation}

where

\begin{equation}
\bm{\mathcal{F}}\left( \bm{X} \right) = \left( p\frac{\beta}{N}S(t)(I(t) + A(t)),\quad (1 - p)\frac{\beta}{N}S(t)(I(t) + A(t)) \right)^{T},\tag{37}\label{eq:37}
\end{equation}

and

\begin{equation}
\bm{\mathcal{V}}\left( \bm{X} \right) = \left( (\gamma + \mu)I(t),\quad \kappa A(t) \right)^{T}.\tag{38}\label{eq:38}
\end{equation}

Now, the next-generation matrix is given by the product
\(\bm{F}\bm{V}^{- 1}\). The matrices \(\bm{F}\) and
\(\bm{V}\) are the Jacobians of
\(\bm{\mathcal{F}}\left( \bm{X} \right)\) and
\(\bm{\mathcal{V}}\left( \bm{X} \right)\) evaluated at the disease-free equilibrium
\(\bm{Y}^{0}\). Both \(\bm{F}\) and \(\bm{V}\) are 2×2
matrices, given by

\begin{equation}
\bm{F} = \begin{pmatrix}
\begin{matrix}
p\beta \\
(1 - p)\beta
\end{matrix} & \begin{matrix}
p\beta \\
(1 - p)\beta
\end{matrix}
\end{pmatrix},\tag{39}\label{eq:39}
\end{equation}

and

\begin{equation}
\ \bm{V} = \ \begin{pmatrix}
\begin{matrix}
\gamma + \mu \\
0
\end{matrix} & \begin{matrix}
0 \\
\kappa
\end{matrix}
\end{pmatrix}.\tag{40}\label{eq:40}
\end{equation}

Apparently, matrix \(\bm{V}\) is invertible, where the inverse
matrix has the form

\begin{equation}
\bm{V}^{-1} = \ \begin{pmatrix}
\begin{matrix}
\frac{1}{\gamma + \mu} \\
0
\end{matrix} & \begin{matrix}
0 \\
\frac{1}{\kappa}
\end{matrix}
\end{pmatrix}.\tag{41}\label{eq:41}
\end{equation}

Finally, \(R_{0}\) is the spectral radius of the next-generation matrix
\(\bm{F}\bm{V}^{- 1}\), leading to the expression

\begin{equation}
R_{0} = \rho\left( \bm{F}\bm{V}^{- 1} \right) = (1 - p)\frac{\beta}{\kappa} + p\frac{\beta}{\gamma + \mu},\tag{42}\label{eq:42}
\end{equation}

where \(\rho(\bm{A})\) stands for the spectral radius of matrix
\(\bm{A}\). Considering that parameters \(\gamma\), \(\kappa\),
\(p\), \(\mu\) and \(R_{0}\) are known, we can determine the appropriate
\(\beta\) value.

\end{document}